\documentclass[nohyper, 11pt,letterpaper]{JHEP3}
\usepackage{epsfig,yfonts}
\def\be{\begin{equation}}
\def\ee{\end{equation}}
\def\ba{\begin{eqnarray}}
\def\ea{\end{eqnarray}}
\def\gw{\textswab{w}}
\def\gq{\textswab{q}}
\def\gR{\textswab{R}}
\def\gF{\textswab{F}}
\def\cG{{\cal G}}

\newcommand{\reef}[1]{(\ref{#1})}

\newcommand{\labell}[1]{\label{#1}} 

\newcommand{\eg}{{\it e.g.,}\ }
\newcommand{\ie}{{\it i.e.,}\ }
\newcommand{\comment}[1]{{\bf [[#1]]}}
\newcommand{\mt}[1]{\textrm{\tiny #1}}
\newcommand{\dd}{\tilde{d}}
\newcommand{\vm}{v_\mt{lim}}
\newcommand{\qm}{\gq_\mt{crit}}

\newcommand{\qo}{{\Omega}}
\newcommand{\qg}{{\Gamma}}
\def\nc {N_\mt{c}}
\def\nf {N_\mt{f}}
\def\gym {g_\mt{YM}}
\newcommand{\td}{T_\mt{dec}}
\newcommand{\tf}{T_\mt{fun }}

\newcommand{\mq}{M_\mt{q}}      
\newcommand{\mbar}{\bar{M}}
\newcommand{\gs}{g_\mt{s}}
\newcommand{\ls}{\ell_\mt{s}}
\newcommand{\ids}{I_\mt{D7}}

\newcommand{\nq}{n_\mt{q}}

\def\sac{\, , \,\,\,\,\,}

\newcommand{\overlrarrow}[1]{\vbox{\ialign{##\cr\cr
                  \leftrightarrowfill\crcr\noalign{\kern-1pt\nointerlineskip}
                  $\hfil\displaystyle{#1}\hfil$\crcr}}}

\newcommand{\muq}{\mu_\mt{q}}

\title{The fast life of holographic mesons}
\author{Robert C. Myers$^{a,b}$ and Aninda Sinha$^a$ \\
$^a$ {\it Perimeter Institute for Theoretical Physics, Waterloo,
Ontario N2L 2Y5, Canada}\\
$^b$ {\it Department of Physics and Astronomy, University of
Waterloo,
Waterloo, Ontario}\\
\ \ {\it N2L 3G1, Canada}\\

\vskip .5cm {\rm E-mail:}\ \ {\tt
rmyers,$\,$asinha@perimeterinstitute.ca}}

\date{\today}

\abstract{We use holographic techniques to study meson
quasiparticles moving through a thermal plasma in ${\cal N}=2$
super-Yang-Mills theory, with gauge group $SU(\nc)$ and coupled to
$\nf$ flavours of fundamental matter. This holographic approach
reliably describes the system at large $\nc$, large 't Hooft
coupling and $\nf/\nc \ll 1$. The meson states are destabilized by
introducing a small quark density $\nq$. Spectral functions are used
to examine the dispersion relations of these quasiparticles. In a
low-momentum regime, the quasiparticles approach a limiting velocity
which can be significantly less than the speed of light. In this
regime, the widths of the quasiparticles also rise dramatically as
their momentum approaches a critical value $\qm$. While the spectral
functions do not display isolated resonances for $\gq>\qm$, the
dispersion relations can be extended into this high-momentum regime
by studying the dual quasinormal modes. A preliminary qualitative
analysis of these modes suggests that the group velocity rises to
the speed of light for $\gq\gg\qm$.}


\preprint{arXiv:0804.2168 [hep-th]}

\begin{document}

\section{Introduction}

A large class of strongly coupled gauge theories can be studied
using the gauge/gravity duality \cite{juan,bigRev}. The
gauge theories that are currently amenable to study with holographic
techniques are very different from real world QCD, \eg current
calculations are restricted to large $\nc$ and large 't Hooft
coupling. However, this approach has still proven to be a fruitful
framework with which to gain new insights into the strongly coupled
quark-gluon plasma -- see, \eg \cite{shear,seat}.

With this aim in mind, holographic techniques have been applied to
study the thermal properties of $\nf$ flavours of fundamental matter
in ${\cal N}=2$ $SU(\nc)$ super-Yang-Mills (SYM) in a quenched
approximation (\ie $\nf\ll\nc$) \cite{johanna,prl,long,recent}. The
gravity dual for this field theory consists of $\nf$ probe D7-branes
in the black hole background generated by $\nc$ D3-branes. In this
system, the fundamental matter generically undergoes a first order
phase transition at some temperature $\tf$. The low-temperature
phase of the theory is described by `Minkowski embeddings' of the
probe branes (see fig.~\ref{embeddings}) in which the branes sit
entirely outside the black hole \cite{prl,long}. In this phase, the
meson spectrum is discrete and exhibits a mass gap. Above the
critical temperature $T_\mt{fun}$, the branes are characterised by
`black hole' embeddings which extend through the event horizon.  In
this phase, the meson spectrum is continuous and gapless
\cite{long,spectre,hoyos}.  Thus, this large-$\nc$, strong coupling
phase transition is associated with the dissociation of the mesons.
It has been suggested that this physics is in qualitative agreement
with that of heavy quarkonium in QCD \cite{long}. Studies from
lattice QCD \cite{lattice} suggest that such mesons survive the
deconfinement phase transition at $\td \sim 175 $ MeV and remain as
relatively well-defined resonances up to temperatures of $2-3\,
\td$.
\FIGURE{
 \includegraphics[width=1 \textwidth]{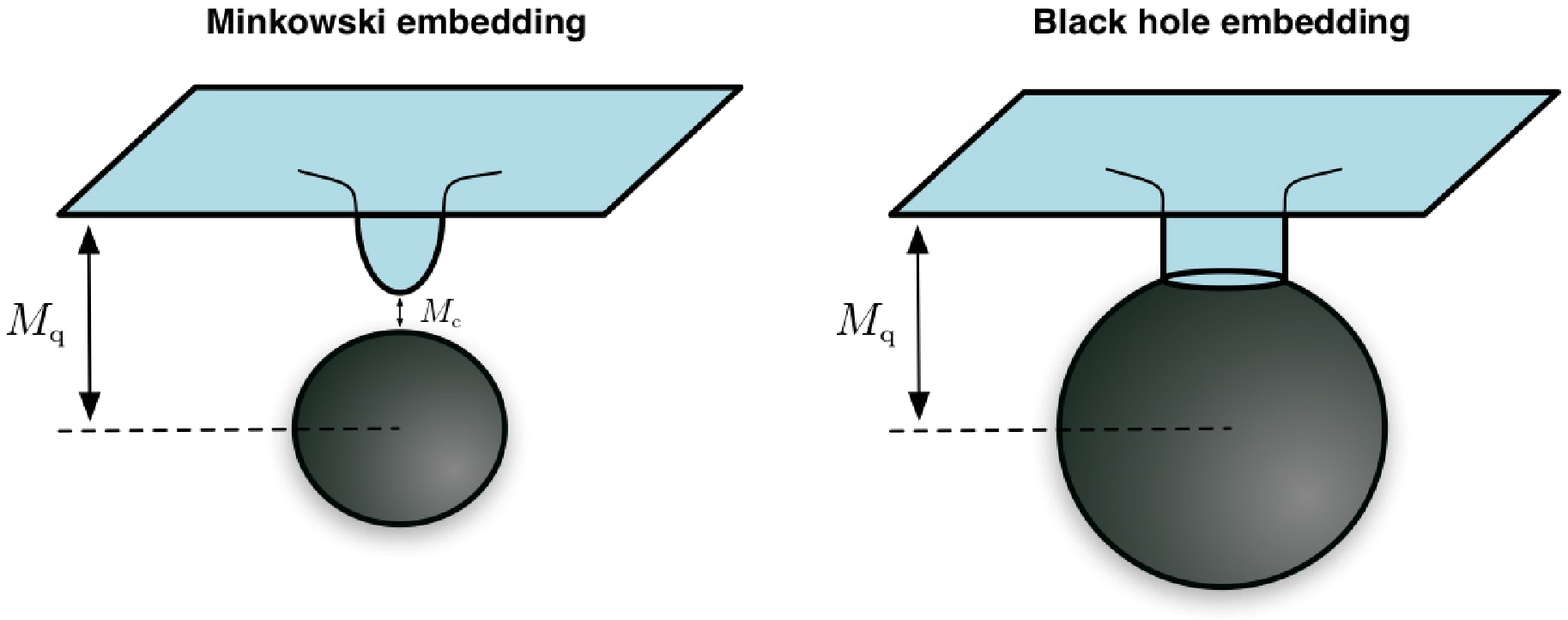}
\caption{Possible embeddings for probe D7-branes in the background
black hole geometry of D3-branes.} \label{embeddings}}

Another interesting feature that was discovered for the holographic
mesons in the low temperature phase is that moving through the
thermal plasma,  at large momentum they approach a limiting
velocity with $\vm<1$ \cite{long,ejaz}. On the gravity side, this
can be understood as the dual excitations travelling at the local
speed of light near the minimum radius reached by the D7-branes.
Hence in the field theory, one finds the velocity
 \be
\vm={dx\over
dt}=\left.\sqrt{-g_{tt}/g_{xx}}\right|_{\rho=\rho_{min}}\labell{speed}
 \ee
Because of the redshift near the black hole horizon, this yields a
result which can be much less than one. As discussed in \cite{ejaz},
this result \reef{speed} can be re-expressed as
 \be
\vm^2\simeq1-{\lambda^2\over4} \left({T\over M_q}\right)^4\ .
\labell{speed2}
 \ee
We must note that the limiting speed does not actually represent a
`speed limit' for the mesons. That is, a careful analysis show that
the group velocity actually approaches $\vm$ from above and so the
maximum group velocity is actually slightly larger than this
asymptotic value \cite{ejaz}. In any event, the effect that $\vm<1$
is somewhat surprising as one's naive intuition would be that a
meson traveling through the thermal plasma would eventually reach
the speed of light if the energy/momentum is increased to
arbitrarily large values. Hence it appears that this new limiting
velocity is a consequence of strong coupling.

As mentioned above in the low temperature phase, the holographic
mesons are stable or rather their widths are suppressed by $1/\nc$.
Of course, this stands in contrast with the heavy-quark mesons
studied in thermal lattice QCD \cite{lattice} which are quite broad
states. In the black hole phase of the holographic model, the meson
excitations are readily absorbed by the black hole horizon. The
quasinormal frequencies of these excitations typically have Im$(
\omega)$ $\sim$ Re$(\omega)$ \cite{hoyos} and so the corresponding
spectral functions do not reveal any quasiparticles \cite{spectre}.
However, black hole embeddings can also arise at low temperatures
when a nonzero quark density $\nq$ is introduced\footnote{Since the
the underlying theory here is supersymmetric, the quark density we
consider arises from a hypermultiplet with both fermions and
bosons.} -- in fact, these are the only physically consistent
embeddings in this situation \cite{findens}. In these backgrounds,
the meson states are again unstable but the width of these states
can be tuned by varying $\nq$ \cite{finitemu,johanna2}. Essentially,
with a small $\nq$, the D7-brane extends down to the horizon with a
narrow neck (as illustrated in fig.~\ref{embeddings2}) and so the
absorption of the meson excitations is limited by the small
effective horizon area of the worldvolume metric. Hence the
corresponding spectral functions display clear quasiparticle states
with narrow widths \cite{johanna2}. In this paper, we begin a study
of the dispersion relations of these quasiparticle states.
\FIGURE{
 \includegraphics[width=0.6 \textwidth]{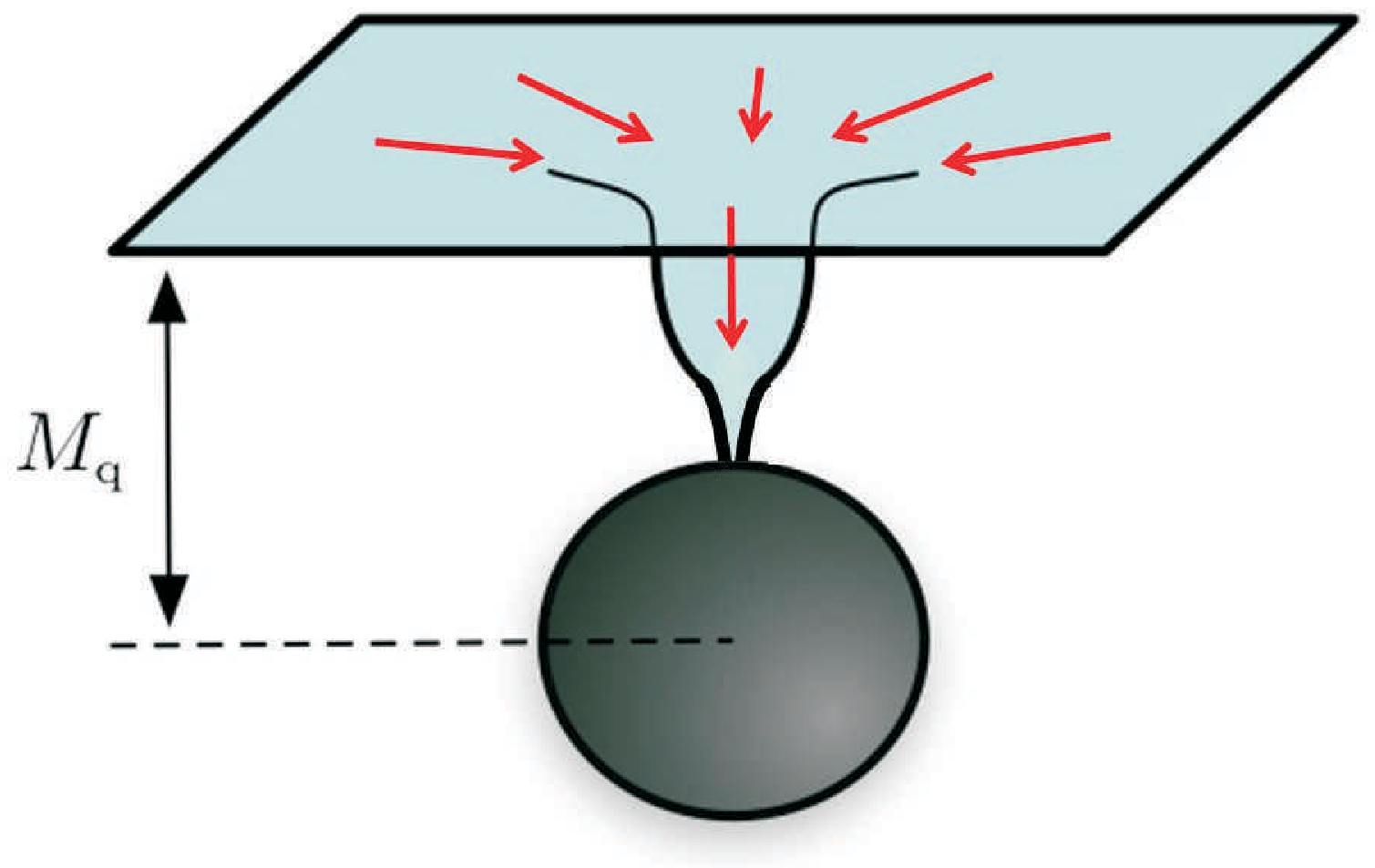}
\caption{D-brane embedding with radial electric field is necessarily
a black hole embedding but the width of the throat is tuneable. }
\label{embeddings2}}

An overview of the paper is as follows: in section \ref{holo} we
introduce the D3/D7-brane framework and review the D7-brane
embeddings and thermodynamics. In section \ref{spectral}, we turn to
the calculation of the spectral function for various meson
operators. By following the position and shape of the quasiparticle
resonances in the spectral function with growing momentum, $\gq$, we
estimate the dispersion relations for the low-lying resonances. With
this approach, the quasiparticles are found to approach the same
limiting velocity found for the case of stable mesons
\cite{long,ejaz}{\footnote{See also \cite{iancu}.}}.  Another interesting phenomena is that the widths
show a dramatic increase as the momentum approaches a critical value
$\qm$. Section \ref{quasi} presents a qualitative discussion of the
quasinormal frequencies using our intuition derived from casting the
relevant radial equation in the form of an effective Schr\"odinger
equation. This framework gives an alternative point of view from
which to understand the effects noted above. Above $\qm$, the
spectral functions do not exhibit isolated quasiparticle resonances
and so in section \ref{beyond}, we use the Schr\"odinger framework
to consider the behaviour of the quasinormal modes in the
high-momentum regime $\gq>\qm$. In previous sections, we defined the limiting velocity $\vm$
  by observing the (real part of the) dispersion relation approached
  a straight line for large $\gq$ but in the regime $\gq<\qm$. Here our
  qualitative analysis suggests that this definition fails in the
  regime $\gq\gg \qm$. Rather the behaviour of the dispersion relation
  changes such that ultimately it approaches an asymptotic slope of
  one in this very high momentum regime.
 In section
\ref{discuss}, we discuss our results and make a few observations
about future directions. Appendix \ref{wonk} provides some details
about a WKB calculation of the quasinormal frequencies using the
Schr\"odinger framework considered in section \ref{quasi}.

\section{Holographic Framework}\label{holo}

Following \cite{prl,long}, we write the background metric for $\nc$
black D3-branes in the decoupling limit as
 \be ds^2 = \frac{1}{2} \left(\frac{u_0 \rho}{L}\right)^2 \left[-
\frac{f^2}{\tilde f} dt^2 + \tilde{f} d\vec{x}^{\,2} 
\right]
 + \frac{L^2}{\rho^2}\left[ d\rho^2 +\rho^2 d\Omega_{\it 5}^2
  \right] \,,
  \labell{geom}
 \ee
where $\rho$ is a dimensionless coordinate and
\be \label{ffff} f(\rho)= 1- \frac{1}{\rho^4} \sac
\tilde{f}(\rho)=1+\frac{1}{\rho^4} \sac L^4 = 4 \pi \gs \nc \ls^4
\,. \ee
This metric possesses a horizon at $\rho=1$ with temperature
 \be T = \frac{u_0}{\pi L^2} \,, \ee
which is identified with the temperature of the dual ${\cal N}=4$
SYM theory. Further the coordinates $\{t,\vec{x} \}$ are identified
with the coordinates of the gauge theory. The string coupling
constant is related to the SYM 't Hooft coupling constant
through\footnote{Note that we are using the standard D-brane
convention here which differs from that of the usual quantum field
theory literature. As explained in appendix D of \cite{long},
$\lambda=\lambda_\mt{QFT}/2$.}
\be \lambda = \gym^2 \nc =  2 \pi \gs \nc \,. \ee
The background D3-brane solution also has a (constant) dilaton and a
Ramond-Ramond field, whose precise form are not needed in the
following.

Introducing $\nf$ D7-branes into the  geometry above corresponds to
coupling $\nf$ fundamental hypermultiplets to the original SYM
theory \cite{flavour}. Before the decoupling limit, the branes are
oriented in the following array:
\begin{equation}
\begin{array}{ccccccccccc}
   & 0 & 1 & 2 & 3 & 4& 5 & 6 & 7 & 8 & 9\\
\mbox{D3:} & \times & \times & \times & \times & & &  &  & & \\
\mbox{D7:} & \times & \times & \times & \times & \times  & \times
& \times & \times &  &   \\
\end{array}
\labell{D3D7}
\end{equation}
This configuration is supersymmetric at zero temperature, which
ensures stability of the system. After the decoupling limit, the D7
branes wrap an $S^3$ inside the $S^5$ of the background geometry.
Adapting the $S^5$ coordinates to this embedding, we write
 \be d\Omega_{\it 5}^2 = d\theta^2 + \sin^2 \theta\, d\Omega_{\it 3}^2
+\cos^2 \theta \,d\phi^2  \ee
in \reef{geom}.  Defining $\chi = \cos \theta$, we describe the
D7-brane embedding as: $\phi=0, \, \chi=\chi(\rho)$.

The derivation of the equations of motion for the D7-brane profile
$\chi(\rho)$ and the gauge field on their worldvolume $A_t$ was
discussed in \cite{findens}. Here we will review a few salient
points and refer the interested reader to \cite{findens} for more
details. The DBI action of the D7-branes may be written
 \be \ids=-\nf T_\mt{D7} \int dt\, d^3\!x \, d\rho \, d\Omega_{\it
3} \frac{\left( u_0 \rho \right)^3}{4} f \tilde{f} (1-\chi^2)
\sqrt{1-\chi^2+\rho^2 \dot{\chi} ^2 - \frac{2
\tilde{f}}{f^2}(1-\chi^2) \dot{\tilde{A_t}}^2} \,, \labell{action}
 \ee
where the dot denotes differentiation with respect to $\rho$ and we
have introduced the dimensionless gauge field $\tilde{A}_t$
\cite{findens}. The asymptotic form is determined by the gauge
field's equation of motion (eqn. $(2.11)$ in \cite{findens}) as
 \be \tilde{A_t} = \frac{2\pi \ls^2}{u_0}\,\muq -
 \frac{\dd}{\rho^2} + \cdots \,, \labell{at}
 \ee
where the constant $\muq$ is the quark chemical potential. The
dimensionless constant $\tilde{d}$ is related to the vacuum
expectation value of the quark number density operator with
 \be \nq = \frac{1}{2^{5/2}}\, \nf \nc \sqrt{\lambda}\, T^3\, \dd\,.
\labell{deff1}  \ee
The equation of motion for $\chi$ (eqn.~$(2.17)$ in \cite{findens})
implies  that the D7-brane profile behaves asymptotically as
 \be
\chi = \frac{m}{\rho}+\frac{c}{\rho^3} + \cdots \,, \labell{asymp}
 \ee
where the dimensionless constants $m$ and $c$ are proportional to
the quark mass and condensate, respectively \cite{prl,long}. In
particular, $m=\mbar/T$ where
\be \mbar = \frac{2\mq}{\sqrt{\lambda}} = \frac{M_\mt{gap}}{2\pi}
\labell{mbar}  \ee
is (up to a factor) the meson mass gap $M_\mt{gap}$ at zero
temperature \cite{us-meson}.

As only $\partial_\rho\tilde{A}_t$ enters the action \reef{action},
$\dd$ is a conserved integral of motion and gauge field equation
(Gauss' law) yields
 \be\label{gauss}
\partial_\rho\tilde{A}_t = 2 \tilde{d} \frac{f
\sqrt{1-\chi^2+\rho^2 \dot{\chi}^2}} {\sqrt{\tilde{f}(1-\chi^2)
[\rho^6 \tilde{f}^3 (1-\chi^2)^3+8 \tilde{d}^2]}}\,. \ee
Substituting this expression into the equation of motion for the
profile $\chi$ then gives
 \be\labell{chieom}
\partial_\rho\left[{1\over\sqrt{\Delta}}{\rho^5 f\tilde
f(1-\chi^2)\dot\chi\over \sqrt{1-\chi^2+\rho^2\dot{\chi}^2}}\right]
= - {1\over\sqrt{\Delta}}{\rho^3 f\tilde f\chi\over
\sqrt{1-\chi^2+\rho^2\dot{\chi}^2}}
\left[3\Delta(1-\chi^2+\rho^2\dot{\chi}^2) -\rho^2 \dot{\chi}^2
\right]\nonumber\,.
 \ee
where $\Delta$ is given by
 \be\label{Delta}
\Delta={\rho^6\tilde f^3(1-\chi^2)^3\over \rho^6\tilde
f^3(1-\chi^2)^3+8\tilde d^2}\,.
 \ee

\subsection{Black hole embeddings}\label{hole}

A key point for the following analysis is that if $\nq \neq 0$ (\ie
$\dd\neq0$) then the only physically consistent embeddings for the
D7-branes are black hole embeddings \cite{findens,finitemu}. Simply
stated, a nonzero density of quarks is dual to a worldvolume
electric field, \ie a nonvanishing $\dot{\tilde{A_t}}$. In turn,
this electric field can be interpreted as a finite number of
(fundamental) strings dissolved in the probe D7-branes. Since these
strings cannot simply terminate, it is not possible for the
D7-branes to close off smoothly above the horizon.

Of course, the defining feature of the black hole embeddings is that
the probe D7-branes reach the event horizon at $\rho=1$. We can only
solve the profile's equation of motion (\ref{chieom}) using
numerical techniques. Generally we integrate out from the horizon
with the following boundary conditions
 \be
\chi(\rho=1)=\chi_0\,,\quad \partial_\rho \chi(\rho=1)=0\,.
 \ee
The series expansion of $\chi$ around $\rho=1$ takes the
form\footnote{In the numerics, we used this expansion to specify the
boundary conditions slightly away from $\rho=1$.}
 \be
\chi=\chi_0-{3 \chi_0 (1-\chi_0^2)^3\over 4(\tilde
d^2+(1-\chi_0^2)^3)}(\rho-1)^2(2-\rho)+O((\rho-1)^4)\,.
 \ee
Thus as $\chi_0$ approaches 1 (from below) with $\tilde d$ fixed,
the profile develops a long narrow throat extending out from
$\rho=1$ -- as illustrated in fig.~\ref{profile}.

The black hole embeddings were so named because the induced geometry
on the D7-brane worldvolume is a black hole geometry. The area of
the induced horizon, which is proportional to $(1-\chi_0^2)^{3/2}$,
controls the lifetime of excitations on the brane \cite{spectre}. As
discussed below, tuning $\chi_0$ close to 1 will allow us to produce
long-lived quasiparticles. However, in the following, we will want
to compare the lifetimes of various quasiparticles while keeping the
quark mass fixed. Recall that the latter is determined from the
asymptotic form of the profile \reef{asymp} at large $\rho$.
Implicitly then with our numerical approach, the constants $m$ and
$c$ in \reef{asymp} are functions of both $\chi_0$ and $\tilde d$.
Hence adding the parameter $\dd$ is the key to allowing us to vary
$\chi_0$ (\ie the quasiparticle lifetimes) while holding $m$ (\ie
the quark mass) fixed.
\FIGURE[ht]{
 \includegraphics[width=0.6 \textwidth]{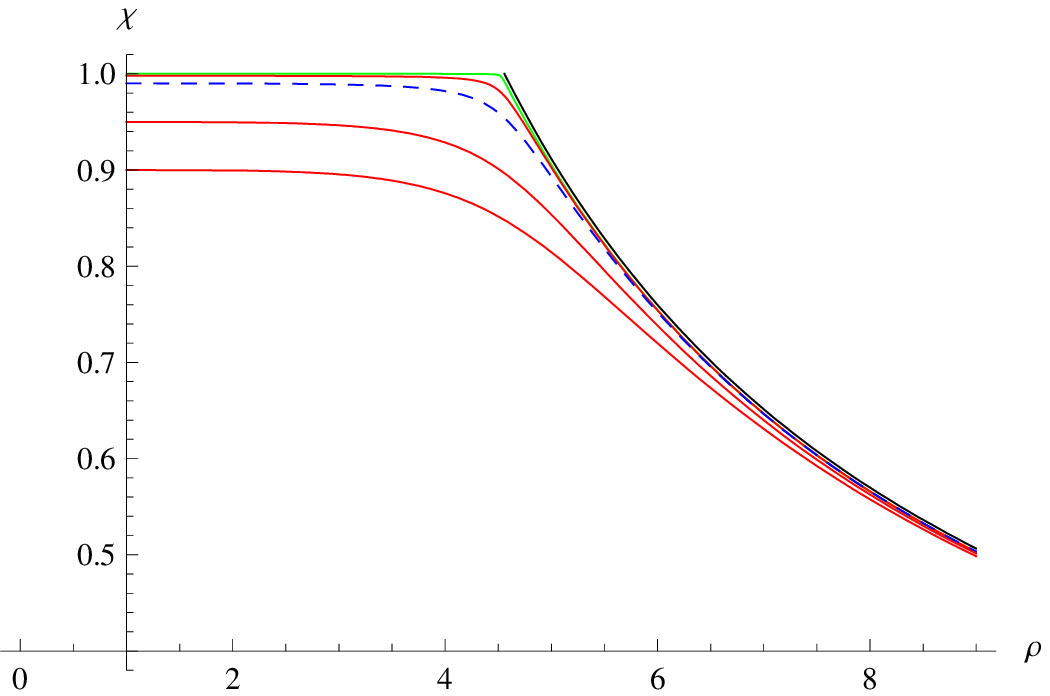}
\caption{The profile $\chi$ as a function of $\rho$ for different
values of $\chi_0, \tilde d$  with fixed $m=4.56$. To keep $m$
fixed, $\tilde d$ is decreased as $\chi_0$ increases . The black
line represents the profile for the corresponding Minkowski
embedding with $\dd=0$. } \label{profile}}

We note that the zero-temperature limit of these configurations is a
subtle one \cite{zero}.

\subsection{Minkowski embeddings}
We comment briefly on Minkowski embeddings with $\dd=0$ with
reference to the limiting velocity discussed in the introduction. To
describe these embeddings, we make the following coordinate
transformation \cite{long}
 \be\label{minkc}
\rho^2=r^2+R^2\,,\quad \chi=R/\rho\,,
 \ee
whereby the equation of motion for the profile $R(r)$ becomes
 \be\label{outout}
\partial_r\left[{r^3f\tilde f\,\partial_r R\over
\sqrt{1+(\partial_r R)^2}} \right] = {8\,r^3 R\over
(r^2+R^2)^5}\sqrt{1+(\partial_r R)^2}
 \ee
where $f$ and $\tilde f$ are defined as in \reef{ffff} but now
expressed in terms of $r$ and $R$ using \reef{minkc}. The boundary
conditions specified at $r=0$ (\ie $\chi=1$) are: $R=R_0,
\partial_r R=0$ with $R_0>1$. Near the axis $r=0$, the embedding has
an expansion.
 \be
R=R_0+{r^2\over R_0(R_0^8-1)}+{5 R_0^{16}+5 R_0^8-3\over 3 R_0^3
(R_0^8-1)^3}\,r^4+O(r^6)\,,
 \ee
which, as expected, clearly illustrates that $r=0$ is the point of
closest approach between the D7-brane and the horizon.  As commented
above, the meson excitations on the Minkowski embeddings were found
to approach a limiting velocity at large momentum \cite{long}. This
velocity is given by
 \be\label{speedlim}
\vm=\left.\sqrt{-{g_{tt}\over g_{xx}}}\right|_{\rho=\rho_{min}} =
\left.{f\over \tilde f}\right|_{r=0}={1-1/R_0^{\,4}\over
1+1/R_0^{\,4}}\,.
 \ee

\section{Meson spectral functions}\label{spectral}

A fruitful application of gauge/gravity duality techniques has
proven to be the study of thermal properties of the strongly coupled
gauge theories through their finite-temperature correlation
functions \cite{new}. In a holographic framework, the spectral
functions are typically easier to compute than the full correlators
and the poles (and associated residues) of the correlators are still
reflected in the corresponding spectral functions. According to the
holographic dictionary, the poles are determined by the quasinormal
spectrum of a dual bulk field fluctuations, whose study is a
technically challenging problem \cite{quasi} --- we begin to address
these calculations in sections \ref{quasi} and \ref{beyond}. In
contrast, the spectral function is given by the imaginary part of
the retarded correlator,
\begin{equation}\label{hockey}
 \gR(\omega, {\bf q} ) = -2\, {\rm Im}\, G^R (\omega, {\bf q})\,,
\end{equation}
and is determined by appealing to standard calculations of bulk
field correlators \cite{Son:2002sd}.

The retarded correlators have poles in the lower
half-plane of complex frequency, which we may assume have the form
 \be\label{tard}
G^R \sim \frac{A}{\omega - \Omega (q,\alpha) + i\Gamma(q,\alpha)}\,,
 \ee
where $\alpha$ represents any additional relevant parameters, \eg
the temperature or quark density. From such a pole, the spectral
function receives a contribution
 \be\label{peek}
\gR (\omega) \sim \frac{2A\Gamma}{(\omega - \Omega)^2 + \Gamma^2}\,.
 \ee
Thus in the vicinity of $\omega = \Omega$, the spectral function has
a peak characterized by a width $\Gamma$. A quasiparticle
interpretation can be given to the peak if it satisfies the Landau
criterion: $\Gamma \ll \Omega$. Hence the masses and lifetimes of
quasiparticles can be extracted from the holographic spectral
functions.

The spectral function $\gR (\omega)$ also has a characteristic form
in the `high-frequency' limit. This behaviour is determined by the
leading short-distance singularity
 \be
\lim_{(t^2-{\bf x}^2)\rightarrow0} \langle {\cal O}(t, {\bf
x})\,{\cal O}(0) \rangle=\frac{{\cal A}}{|t^2-{\bf
x}^2|^\Delta}+\cdots \, ,\labell{sing}
 \ee
where $\Delta$ denotes the dimension of the operator $\cal O$ and
$\cal A$ is a dimensionless constant. A Fourier transform then leads
to the following contribution to the spectral function
 \be
\gR (\omega)\sim  {\cal A}\, (\omega^2-q^2)^{\Delta-2}\,
.\labell{hightail}
 \ee
Note that this high-frequency tail is Lorentz invariant and shows no
indications of a limiting velocity. This should be expected as it
describes the very high-energy/short-distance behaviour, which is
independent of temperature \cite{spectre}.

In the present study on the gravity side, the D7-brane embeddings
extend through the event horizon of the AdS$_5$ black hole, which
describes the theory at finite temperature. As indicated by the name
`black hole embedding', the metric induced on the worldvolume of the
D7-branes is itself a black hole. Even though the latter geometry
does not obey Einstein's equations, the analysis of the hydrodynamic
physics found previously for bulk fields, \eg \cite{new}, is readily
transferred to the worldvolume fields on the D7-brane \cite{spectre}
and hence we can examine the spectral function for various mesonic
operators following the techniques introduced in \cite{techn}.

Here, the area of the horizon induced in the worldvolume geometry
controls how quickly excitations on the D7-branes are absorbed by
the black hole. This absorption rate then determines the lifetime or
width of the corresponding quasiparticles in the dual gauge theory.
For a black hole embedding with a large horizon area (\eg the
typical situation in the high temperature phase when $\dd=0$), one
expects to find that $\Gamma\sim\Omega$ \cite{hoyos}. Further the
corresponding spectral functions are essentially featureless beyond
exhibiting the characteristic high-frequency tail \reef{hightail}.
However, following the discussion in the previous section, by tuning
$\chi_0$ close to 1, the horizon area and consequently the
quasiparticle widths shrink. Hence this tuning can bring us into a
regime where the spectral functions display distinct peaks with
$\Gamma\ll\Omega$ \cite{johanna2}. We re-iterate that having the
freedom to independently tune both $\dd$ and $\chi_0$ is the key to
producing these small widths while holding $m$ (\ie the quark mass)
fixed.

In the following, we calculate the spectral densities for the
flavour current $J^\mu$ which is dual to the worldvolume gauge field
$A_\mu$ on the supergravity side \cite{findens}. These calculations
are a simple extension to finite spatial momentum of the spectral
function calculations appearing in \cite{johanna2}. We begin by
writing the full gauge field as\footnote{Implicitly we are working
with the dimensionless gauge field scaled as in \cite{findens}.}
 \be
\hat A_\mu(\rho,{\bf x})=\delta^t_\mu\, \tilde
A_t(\rho)+A_\mu(\rho,{\bf x})\,.
 \ee
Here $\tilde A_t$ denotes the background gauge field \reef{gauss}
producing the finite quark density, while $A_\mu$ denotes a
fluctuation. For simplicity, we assume here that $A_\mu$ only
depends on $\rho$ and the Minkowski coordinates (and not the
internal coordinates on the $S^3$). To determine the linearized
equations of motion for the fluctuations, we expand the DBI action
to quadratic order. The resulting gauge field lagrangian is
 \be\label{simple}
{\cal L}=-{1\over
4}\sqrt{|\cG|}\left[\cG^{\mu\alpha}\cG^{\beta\gamma}F_{\alpha\beta}
F_{\gamma\mu}-{1\over
2}\cG^{\mu\nu}\cG^{\sigma\gamma}F_{\mu\nu}F_{\sigma\gamma}\right]\,,
 \ee
where $\cG=g+\tilde F$ being the sum of the background metric and
background electromagnetic field. Note that the last term does not
vanish since $\cG$ is no longer a symmetric matrix in general -- see
comments below, however. The linearized equations of motion for $A$
are then
 \be
\partial_\nu\left(\sqrt{| \cG|}\left[\cG^{\nu\mu}\cG^{\sigma\gamma}
+{1\over2}\cG^{[\nu\sigma]}\cG^{\mu\gamma}
\right]F_{\mu\gamma}\right)=0\,.
 \ee
Next we consider the Fourier transform of $A_\mu$
 \be
A_\mu(\rho,{\bf x})=\int {d^4k\over (2\pi)^4}e^{i{\bf k}\cdot {\bf
x}}A_{\mu}(\rho,{\bf k})\,.
 \ee
If we restrict the momentum vector to $(\omega,q,0,0)$, the
components of the electric field are given by
\be E_x=\omega A_x+q A_0\,,\qquad E_{y,z}=\omega A_{y,z}\,.\ee

In the analysis which follows in this section, we will focus on the
transverse fields $E_{y,z}$ both of which we will denote by $E_T$. The
analysis for the longitudinal component $E_x$ is more involved on
two counts. First, it is only for this mode that the second term in
the lagrangian \reef{simple} contributes to the equation of motion.
The second complication is that, in fact, the quadratic lagrangian
above is inadequate to describe these modes. This is because with
the background field $\tilde F_{\rho t}$, these longitudinal modes
mix with the scalar fluctuations $\delta\theta$. Hence we leave the
complete analysis of these coupled modes for the
future.\footnote{These complications are not evident from the analysis in
\cite{johanna2} due to
simplifications at $\gq=0$, for e.g., $\delta\theta$ only mixes with $A_0$.}

The linearized equation of motion for $E_T$ can be written as
 \be\labell{ETeom}
\partial_\rho (F\partial_\rho E_T)+G\left(\gw^2  {\tilde f^2\over
\Delta f^2}-\gq^2 \right)E_T=0
 \ee
with
 \be\label{core}
F={\rho^3 f\over \sqrt{\Delta}} {(1-\chi^2)^2\over \sqrt{1-\chi^2+\rho^2\chi'^2}}\,,\quad
G={8\sqrt{\Delta}\over \rho}{f\over \tilde
f}\sqrt{1-\chi^2+\rho^2\chi'^2}(1-\chi^2)\,.
 \ee
Here we have also introduced $\gw\equiv\omega/(2\pi T)$ and
$\gq\equiv q/(2\pi T)$. Further recall that $\Delta$ is defined in
\reef{Delta} and goes to unity when $\tilde d=0$. The retarded Green
function for $E_T$ is given by
 \be
G^R={N_f N_c T^2\over 8}\left[ {F\over\omega^2}{\partial_\rho
E_T\over E_T} \right]_{\rho\rightarrow\infty}\,.
 \ee
Now the spectral function is defined as
 \be
\gR(\omega,k)=-2\,\omega^2\, {\rm Im} G^R\,,
 \ee
where the extra factor of $\omega^2$ is introduced to produce the
spectral function for $A_{y,z}$ -- compare to \reef{hockey}.
Defining
 \be\label{speck}
\gF={F\over \gw^2}{\partial_\rho E_T \over E_T}\,,
 \ee
it follows from (\ref{ETeom}) that
 \be\labell{FTeom}
\partial_\rho\gF+{\gw^2 \over F }\gF^2 +G\left( {\tilde f^2\over \Delta f^2}-{\gq^2\over
\gw^2}\right)=0\,.
 \ee
Thus we have
 \be\label{speck2}
\gR(\omega,q)=-{N_f N_c T^2\over4}\, \gw^2\, {\rm Im}
\gF(\rho\rightarrow\infty)\,.
 \ee
We solved the above equation numerically taking
$E_T(\rho)=(\rho-1)^{-i\gw}e(\rho)$ where $e(\rho)$ is regular at
the horizon with the boundary conditions: $e(1)=1,\ \partial_\rho
e(1)=i\gw/2$. This implies $\gF(1)=-4i\sqrt{(1-\chi_0^2)^3+{\tilde
d^2}}/\gw$.

\FIGURE{
 \includegraphics[width=0.9 \textwidth]{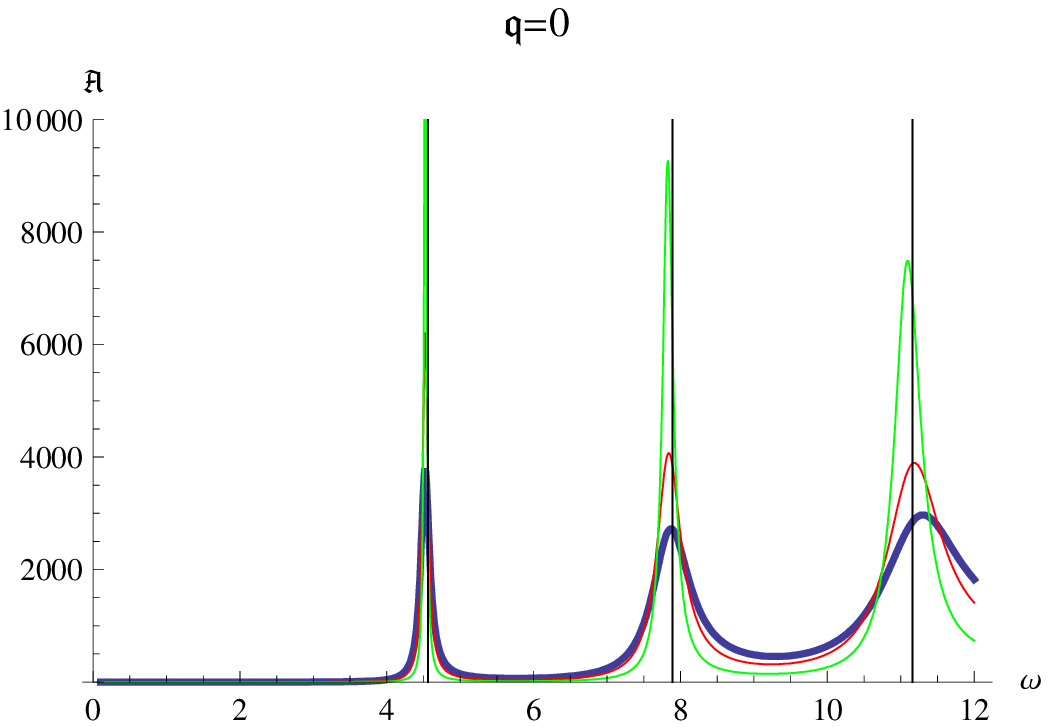}
\caption{ Spectral functions for $m=4.56$ and $\gq=0$. The vertical
black lines correspond the $\delta$-functions appearing for the
Minkowski embedding. The latter appear above at
$\gw=4.56,\,7.89,\,11.16$, respectively. The green line is for
$\tilde d=0.06,\chi_0=0.99755$, red for $\tilde
d=0.15,\chi_0=0.99394$ and blue for $\tilde d=0.25,\chi_0=0.99$. }
\label{dplots}}
Fig.~\ref{dplots} illustrates\footnote{In all our plots in the
following, the spectral function $\gR$ is in units of ${N_f N_c
T^2\over 4}$.} the behaviour of the spectral functions at $\gq=0$ --
this is reproducing the results given in \cite{johanna2}. As
discussed above, to construct the background D7-brane embedding, we
are varying the variables $\chi_0$ and $\dd$ while keeping fixed the
asymptotic mass (or $m$). For the cases illustrated in
fig.~\ref{dplots}, we fixed $m=4.56$ (to within about 0.25\%
accuracy). These examples explicitly show that as the quark density
or induced horizon area increases, the widths of the quasiparticles
increases while their positions remains essentially fixed. The
figure also shows the positions of the corresponding mesons on the
Minkowski embedding with $m=4.56$ (and $\dd=0$) -- these masses were
determined numerically, as described in \cite{long}.

\subsection{Dispersion Relations}\label{disperse}

Now to study the dispersion relations, we consider the spectral
functions with finite spatial momentum $\gq$. We focus our attention
on the first few peaks as these lowest lying resonances are the most
prominent at $\gq=0$. At least for a certain range of $\gq$, the
spectral functions allow us to reconstruct both $\Omega(\gq)$ and
$\Gamma(\gq)$ for these quasiparticles. As might be expected, we will
find that beyond a certain $\qm$, the corresponding poles
\reef{tard} have moved too far away from the real axis to allow us
to identify individual resonances in the spectral functions. We will
also be able to estimate the residue $A(\gq)$ in the regime
$\gq<\qm$.

Fig.~\ref{spectralgrid} illustrates the typical behaviour of a
spectral function as the momentum is increased. In general, one can
observe that the quasiparticle peaks are moving to larger values of
$\gw$ as $\gq$ increases. Similarly, one would say that increasing
$\gq$ causes the individual peaks to become diminished while the
background increases. In particular, the high frequency tail
\reef{hightail} in the present case is
 \be
\gR (\gw)=  4\pi\, (\gw^2-\gq^2)\, \labell{hightail2}
 \ee
with $\Delta=3$ \cite{spectre}. In the fig.~\ref{spectralgrid}, we
see that fairly quickly as the momentum is increased, the spectral
function becomes well modeled by this tail alone. In fact in the
last two plots with $\gq=100$ and 140, the rise in the spectral
function cannot be distinguished from this tail at the scale used
there.
\FIGURE{
 \includegraphics[width=0.9 \textwidth]{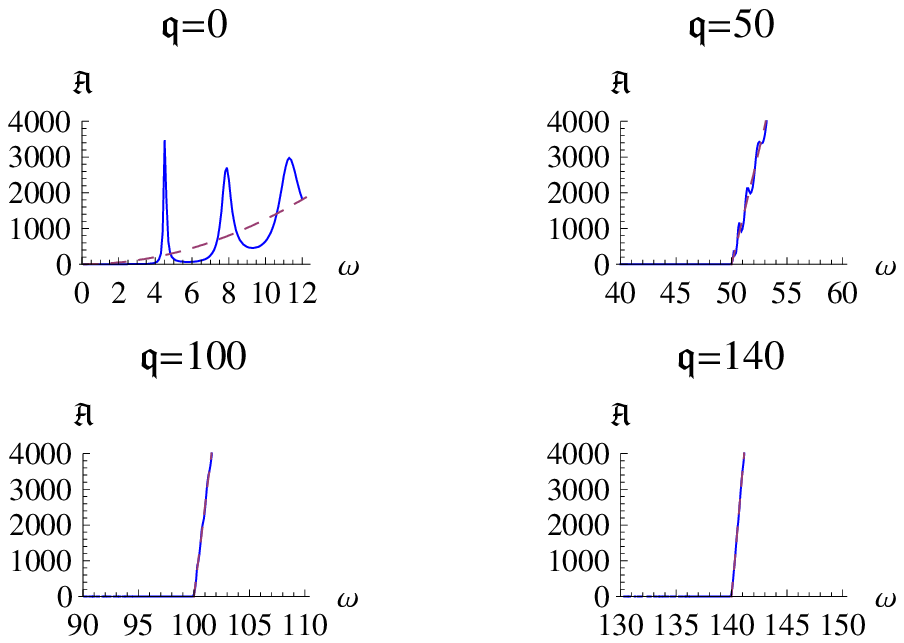}
\caption{Spectral function versus $\gw$ with $\tilde
d=0.25,\chi_0=0.99,m=4.56$ and increasing $\gq$, as indicated. There
are no visible peaks above $\gq=100$. In each plot, the dashed grey
line indicates the high frequency tail \reef{hightail2} for
$\gw\ge\gq$.} \label{spectralgrid}}

\noindent\underline{\bf Energies, $\Omega(\gq)$}: To make our
discussion more quantitative, we followed the positions $\qo_i(\gq)$
of the (first few) peaks as $\gq$ increases for three different
limiting velocities. The three cases are summarized in the following
table:
\be\label{table}
\begin{tabular}{|c|c|c|c|c|c|c|} \hline
   & $\vm$ & $m$ & $\dd$ & $\chi_0$ & $\qo_0$ & $\qg_0$ \\
\hline
I  & 0.995& 4.56& 0.25 & 0.99 & 4.52 & 0.072 \\
II  & 0.651& 1.50 & 0.0005 & 0.9995 & 1.37 & 0.002 \\
III  & 0.343& 1.32 & 0.0001 & 0.99975& 1.03& 0.001 \\
\hline
\end{tabular}
\ee
The columns labelled $\qo_0$ and $\qg_0$ in the table above indicate
the position and width of the first peak at $\gq=0$, respectively,
for each case. In each of these cases, we chose $\qg_0/\qo_0\ll1$ as
this allowed us to follow the first few peaks to much higher
momenta. As we will see below, if one begins with broader peaks at
$\gq=0$ then they dissipate more quickly as $\gq$ is increased. As
described above at \reef{peek}, the positions of such the
peaks\footnote{In the following, the subscript is $i=1,2,3$ to
denote the first three resonances in the spectral function. Further
we identify the position of the resonance with the condition
$\partial_\gw\gR=0$ here.} $\qo_i(\gq)$ should correspond to the
real part of the position of a pole in the corresponding retarded
correlator.\footnote{In the following discussion, we denote the
position of the poles in the complex $\gw$-plane as $\qo-i\qg$.
Hence this usage differs from that in \reef{tard} or \reef{peek} by
a factor of $2\pi T$.}
\FIGURE[ht]{\begin{tabular}{cc}
\includegraphics[width=0.45 \textwidth]{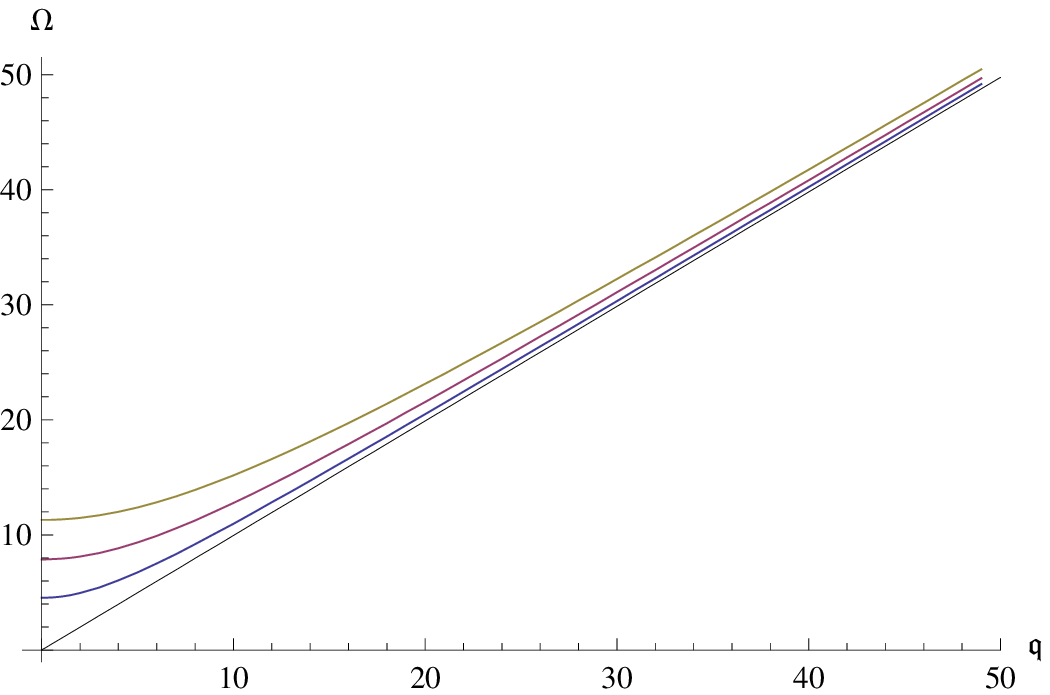}&
\includegraphics[width=0.45 \textwidth]{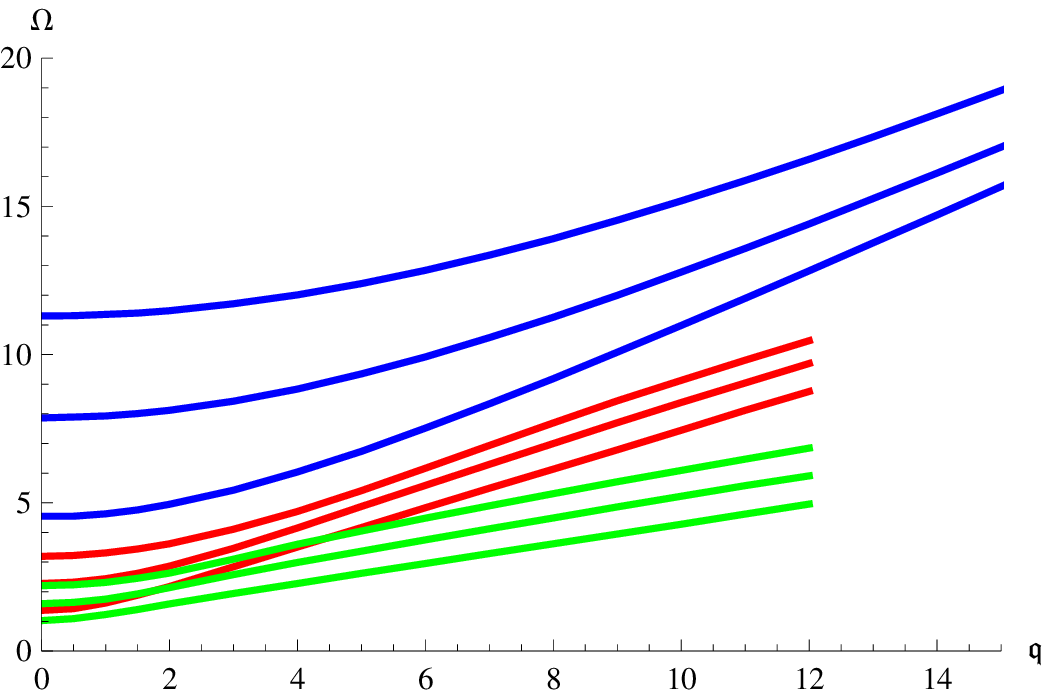}\end{tabular}
\caption{The plot on the left shows the positions of the first three
peaks for case I (see table \reef{table}) as a function of $\gq$.
Within our accuracy, these curves all asymptote to roughly
$\gw=0.995\gq$, which corresponds to the black line. The plot on the
right displays similar dispersion curves for all three cases. The
blue, red and green lines correspond to cases I, II and III,
respectively.} \label{omegavsqwithstrtline}}

The resulting plots of $\qo_i(\gq)$ for these
three examples are displayed in fig.~\ref{omegavsqwithstrtline}.
Note that fig.~\ref{omegavsqwithstrtline} contains two plots which
differ primarily in the horizontal scale for $\gq$ because while the
peaks could be followed out to $\gq\approx100$ for case I
($\vm=.995$), they disappeared around $\gq\approx10$ in the last two
cases -- see discussion below. In all three cases, the curves
$\qo_i(\gq)$ appear asymptote to the straight-line form:
 \be \qo_i(\gq)=\vm\,\gq+a_i+O(1/\gq)\label{strait}
 \ee
found analytically for the Minkowski embeddings in \cite{ejaz}.
Further, as is implicit in our notation, the asymptotic value of the
slope $\partial_\gq\qo_i$ matches very well the limiting velocity
$\vm$ calculated from the corresponding Minkowski embedding with the
same value of $m$. For $\vm=0.995$, this is best illustrated in the
left panel of fig.~\ref{omegavsqwithstrtline}. Fig.~\ref{omegaslope}
shows the first peak for the three different cases fit with a
straight asymptote with slope $\vm$, as given in \reef{table}. This
figure makes clear that the two cases with slower limiting
velocities approach their asymptotic behaviour \reef{strait} more
quickly. We should comment that while the limiting velocities in
\reef{table} gave a good fit to the asymptotic behaviour of these
curves, the accuracy of our numerical calculations was limited. The numerical error is estimated as follows: Using $\vm\approx {R_0^4-1\over R_0^4+1}$, we have
 $ \Delta \vm= 8\Delta R_0 {R_0^3\over R_0^8-1} \vm$. Using $\Delta R_0=\pm 0.01$, this gives $\Delta \vm \approx \pm 0.00004,\pm 0.008, \pm 0.01$ for $\vm=0.995, 0.651, 0.343$ respectively. Hence we are being slightly extravagant in quoting three
  significant figures for the latter two cases.
 We also note that we found that the
constants $a_i$ did not seem to match well with the analytic
expressions given in \cite{ejaz} for the Minkowski embeddings.

\FIGURE[ht] {\includegraphics[width=0.7 \textwidth]{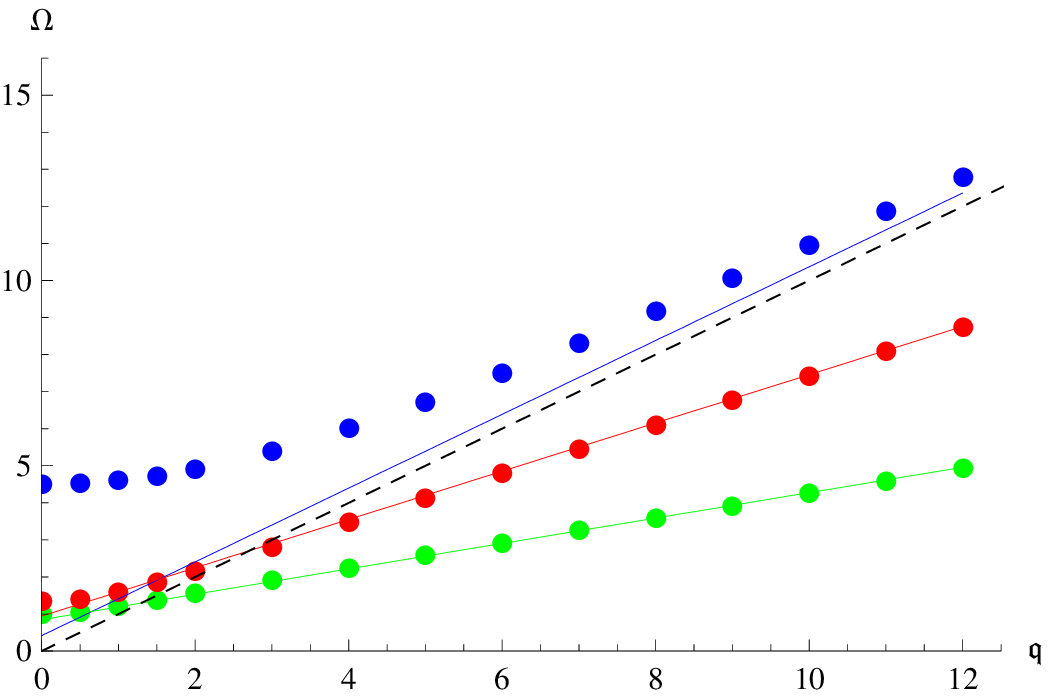}
\caption{The points above correspond to positions of the first peak
from the spectral function for each of the cases in table
\reef{table}. The corresponding straight lines, which match the
asymptotic behaviour in each case, have slopes $\vm=0.995,\ 0.651$
and $0.343$, respectively. The dashed black line has slope one,
corresponding to null four-momenta: $\Omega^2-\gq^2=0$.}\label{omegaslope}}

\noindent\underline{\bf Widths, $\Gamma(\gq)$}: Next we wish to
examine how the widths of the quasiparticles evolve with increasing
momentum. For this purpose, we list four new trial cases in the
table below. Note that the initial widths $\Gamma_0$ are tuned to be
roughly equal in cases IV, V and VI, which were also chosen to match
the same values of $\vm$ as appeared in cases I, II and III in table
\reef{table}. We have also chosen case VII with $\vm=0.651$ but
tuned so that the ratio $\Gamma_0/\Omega_0$ is roughly the same as
in case IV with $\vm=0.995$.
\be\label{table2}
\begin{tabular}{|c|c|c|c|c|c|c|} \hline
   & $\vm$ & $m$ & $\dd$ & $\chi_0$ & $\qo_0$ & $\qg_0$ \\
\hline
IV  & 0.995& 4.56& 0.15 & 0.99394 & 4.52 & 0.044 \\
V  & 0.651& 1.50 & 0.01 & 0.99 & 1.37 & 0.04 \\
VI  & 0.343& 1.32 & 0.0033 & 0.991& 1.03& 0.042 \\
VII  & 0.651& 1.50& 0.004 & 0.996 & 1.37 & 0.015 \\
\hline
\end{tabular}
\ee

As described at \reef{peek}, near the location of an isolated peak
in the spectral function, we expect that the spectral function can
be approximated as
 \be\label{peek2}
\gR\approx {2A\qg\over (\gw-\qo)^2+\qg^2}\,.
 \ee
From this form, we can derive the following expression for the width
of the peak (\ie the imaginary part in the position of the
corresponding pole):
 \be\label{form1}
\Gamma\approx \left.\sqrt{-2{\gR\over \gR''}}\right|_{\gw=\qo}\,,
 \ee
where $'$ denotes differentiation with respect to $\gw$. In the
above expression, we define $\gw=\qo$ as the point where
$\partial_\gw\gR=0$. The results for the width of the first peak as
given by this formula are shown in fig.~\ref{Gammavsqfromplots} (for
each of the four cases in table \reef{table2}). Note that each of
these curves shows a dramatic increase in $\Gamma$ as $\gq$
increases from zero. This gives the first hint that we should find a
certain maximum value of $\gq$ beyond which the quasiparticles do
not exist. Certainly this is observed, \eg in
fig.~\ref{spectralgrid} where the peaks in the spectral function are
simply washed out at large $\gq$.

\FIGURE[ht]{
 \includegraphics[width=0.5 \textwidth]{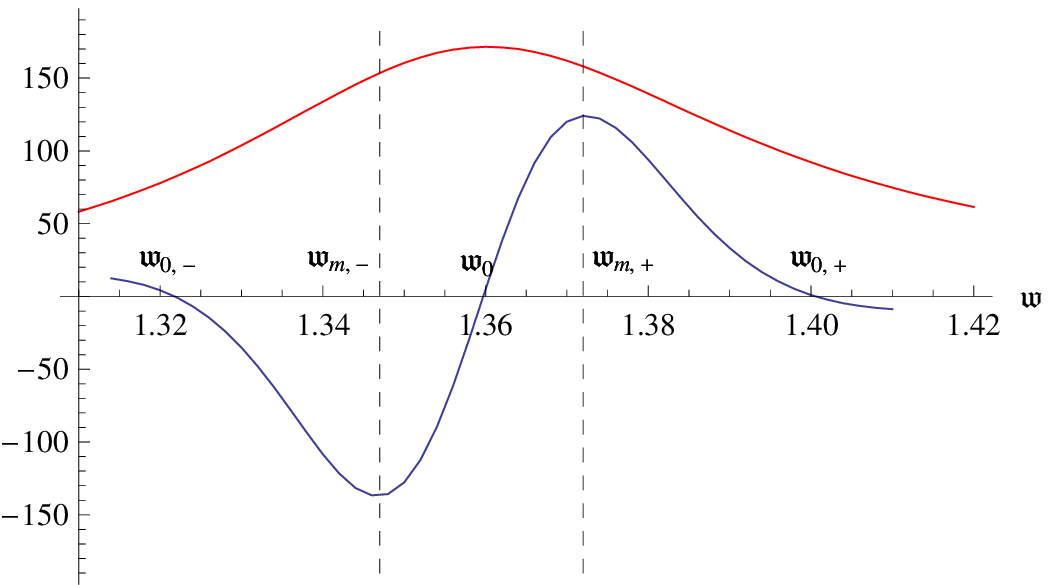}
\caption{A plot of the first peak in $\gR$ (red) and the
corresponding $\partial^3_\gw\gR$ (blue) for case V with $\gq=0$.
This illustrates the positions of the frequencies $\gw_0$,
$\gw_{0,\pm}$ and $\gw_{m,\pm}$ discussed in the text.
$\partial^3_\gw\gR$ has be scaled by 1/200 in this plot.}
\label{neat}}
\FIGURE[ht]{
 \includegraphics[angle=-90,width=1 \textwidth]{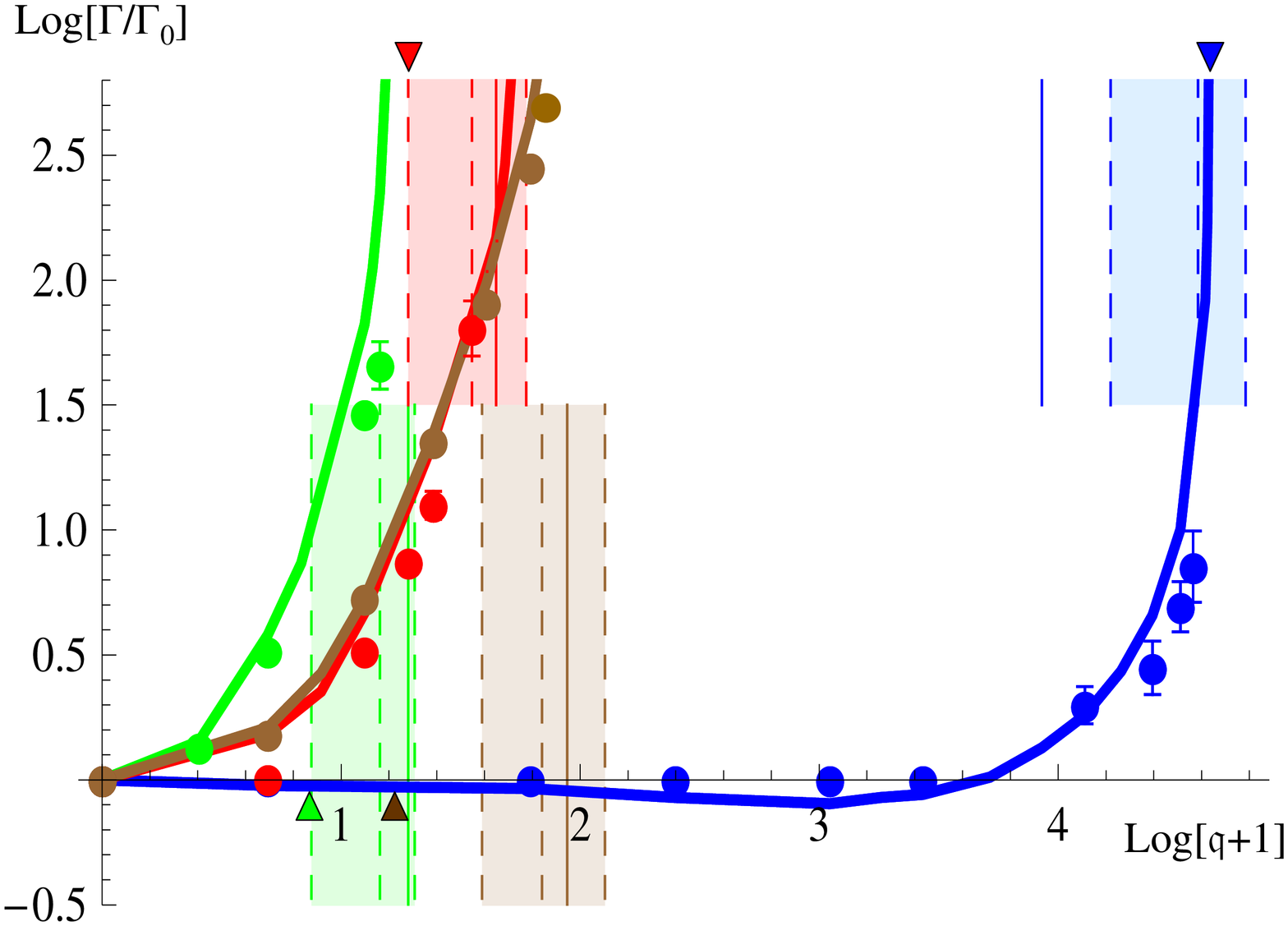}
\caption{$\log(\Gamma/\Gamma_{\gq=0})$ vs $\log(\gq+1)$. The
continuous curves were obtained using the spectral plots and formula
(\ref{form1}). The blue, red, green and brown lines correspond to
cases IV, V, VI and VII in table \reef{table2}, respectively. The
points with `error bars' were obtained using (\ref{next2}). The
vertical dashed lines within the shaded regions indicate the
location where the zero $\gw_{m,+}$ in $\partial_\gw^3\gR$ is
lifted. The left extreme of the shaded regions correspond to $\gq$
when $\gw_{0+}$ is lifted while the right extreme corresponds to
$\gq$ when $\gw_0$ is lifted. The solid vertical lines indicate
where a point of inflexion occurs in the effective potential, as
described in section \ref{quasi}. The arrows indicate where
$\gq=\Omega$.} \label{Gammavsqfromplots}}
Note however, that if the spectral function contains a significant
background contribution near $\gw=\qo$, then the above formula
\reef{form1} will tend to overestimate the true value of $\Gamma$.
In particular, we can seek to improve the approximate form of the
spectral function \reef{peek2} in the vicinity of one of the peaks
by also including the high frequency tail \reef{hightail2}
 \be\label{peek3}
\gR\approx {2A\qg\over (\gw-\qo)^2+\qg^2}+4\pi (\gw^2-\gq^2)\,.
 \ee
Now we observe, however, that by taking the third derivative,
$\partial_\gw^3\gR$, the background introduced by the tail in the
above expression will be eliminated. Further the resulting
expression will have zeros at $\gw_0\equiv\qo$ and
$\gw_{0,\pm}\equiv\qo\pm \qg$ and so in principle, the zeros of
$\partial_\gw^3\gR$ can be used to determine $\qg$. However, rather
than working directly with these zeros, we note that there are also
a maximum and a minimum to either side of the central zero at $\qo$.
A straightforward calculation shows that these extrema occur at
 \be\label{next1}
\gw_{m,\pm}\equiv\qo\pm\delta\ \qg \qquad{\rm with}\ \
\delta=\sqrt{1-{2\over\sqrt{5}}}\simeq 0.3249\,.
 \ee
The above features are illustrated in fig.~\ref{neat}. We can use
these expressions to estimate both the width of a given peak but
also an error because our form \reef{peek2} is not perfect. In
particular, we have
 \be\label{next2}
\bar\Gamma={\gw_{m,+}-\gw_{m,-}\over2\delta}\qquad\Delta\qg={1\over2\delta}
\left(\gw_{m,+}+\gw_{m,-}-2\gw_0\right)\,.
 \ee
The results for the width using these expressions are also shown in
fig.~\ref{Gammavsqfromplots}. One can note that again there is
dramatic rise in $\Gamma$ as $\gq$ increases, again hinting at a
$\qm$. These results for $\bar\Gamma$ seem to agree fairly well with
those originally derived with \reef{form1}. However, this agreement
is not as good for larger values of $\gq$ where the quasiparticle
peaks are washed out and $\Delta\Gamma$ is larger. In particular, as
might be expected, \reef{form1} tends to give a larger width than
\reef{next2} in this regime.

In fact, the increase in the widths is only one of three effects
leading to the dissolution of the quasiparticle peaks in the
spectral functions. The second effect which we consider here is the
decreasing separation of neighbouring peaks or poles along the real
axis. As shown in fig.~\ref{Gammavsqfromplots}, one finds that
$\Delta\Gamma$ from \reef{next2} also increases with increasing
$\gq$. This is indicating a greater asymmetry in $\partial_\gw^3\gR$
around  the position of the peak $\gw_0\equiv\Omega$. Of course,
this is actually indicates that the form \reef{peek3} is becoming a
poor approximation for the spectral function in the vicinity of the
peaks. Essentially the problem is that assuming that the peaks are
isolated is incorrect in this regime. To illustrate this behaviour
for case VI, we show the ratio of the separation of the first two
peaks to the sum of their widths in fig.~\ref{squash}. From our
plot, we see that this ratio approaches one indicating that this
first peak is no longer isolated from the other quasiparticle peaks
at higher $\gw$. The same behavior is observed for the other cases.
For cases IV, V, VII we have for $\gq=0$, the ratio $\log(\Omega_2
-\Omega_1)/(\bar\Gamma_2 +\bar\Gamma_1) =2.54,1.54,2.36$
respectively while the ratio near $\gq$ where $\gw_{m,+}$ becomes a
point of inflection becomes $.28 (\gq=90), .40 (\gq=3.6), .36
(\gq=5)$ in the respective cases.
\FIGURE[hb]{
 \includegraphics[width=0.7 \textwidth]{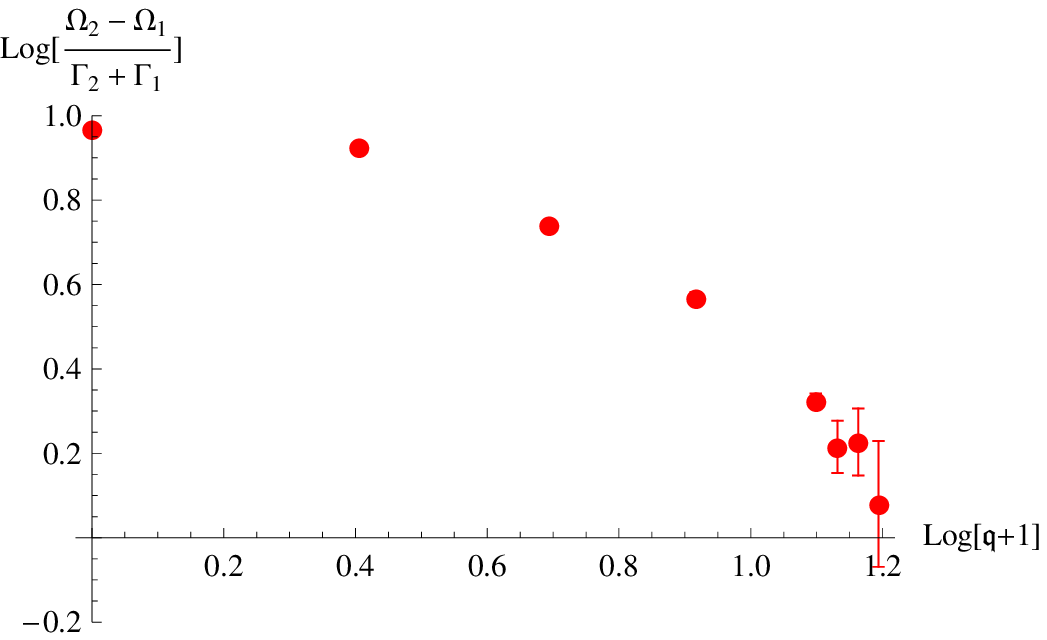}
\caption{Plot of $\log(\Omega_2-\Omega_1)/(\Gamma_2+\Gamma_1)$ for
case VI. The central values arise from using
$\Gamma_{1,2}=\bar\Gamma_{1,2}$ while the upper and lower `error'
bars are determined with $\Gamma_{1,2} = \bar\Gamma_{1,2} \pm
\Delta\Gamma_{1,2}$. The ratio approaches one for larger $\gq$
indicating that these resonances in the spectral function should not
be considered to be isolated peaks.} \label{squash}}

The first peak which we are following then loses its shape due to
the encroachment of the neighbouring peaks. The typical development
as $\gq$ increases is that first the zero at $\gw_{0,+}$ is lifted.
Next the maximum at $\gw_{m,+}$ becomes a point of inflection and
then is subsequently lost. Finally, the zero at $\gw_0$ itself
lifted when it collides with $\gw_{0,-}$ (and simultaneously with
$\gw_{m,-}$). The region where the first peak is losing its shape in
this way is indicated with the shading in
fig.~\ref{Gammavsqfromplots} for each of the four cases. This
dissolution of the first peak is illustrated in detail for case V in
fig.\ref{newone}. As shown, first $\gw_{0,+}$ is lifted around
$\gq=2.6$. This is followed by the maximum $\gw_{m,+}$ becoming a
point of inflexion around $\gq=3.7$ and finally $\gw_0$ is itself
lifted around $\gq=4.9$.
\begin{center}
\FIGURE{
\includegraphics[angle=-90, width=1 \textwidth]{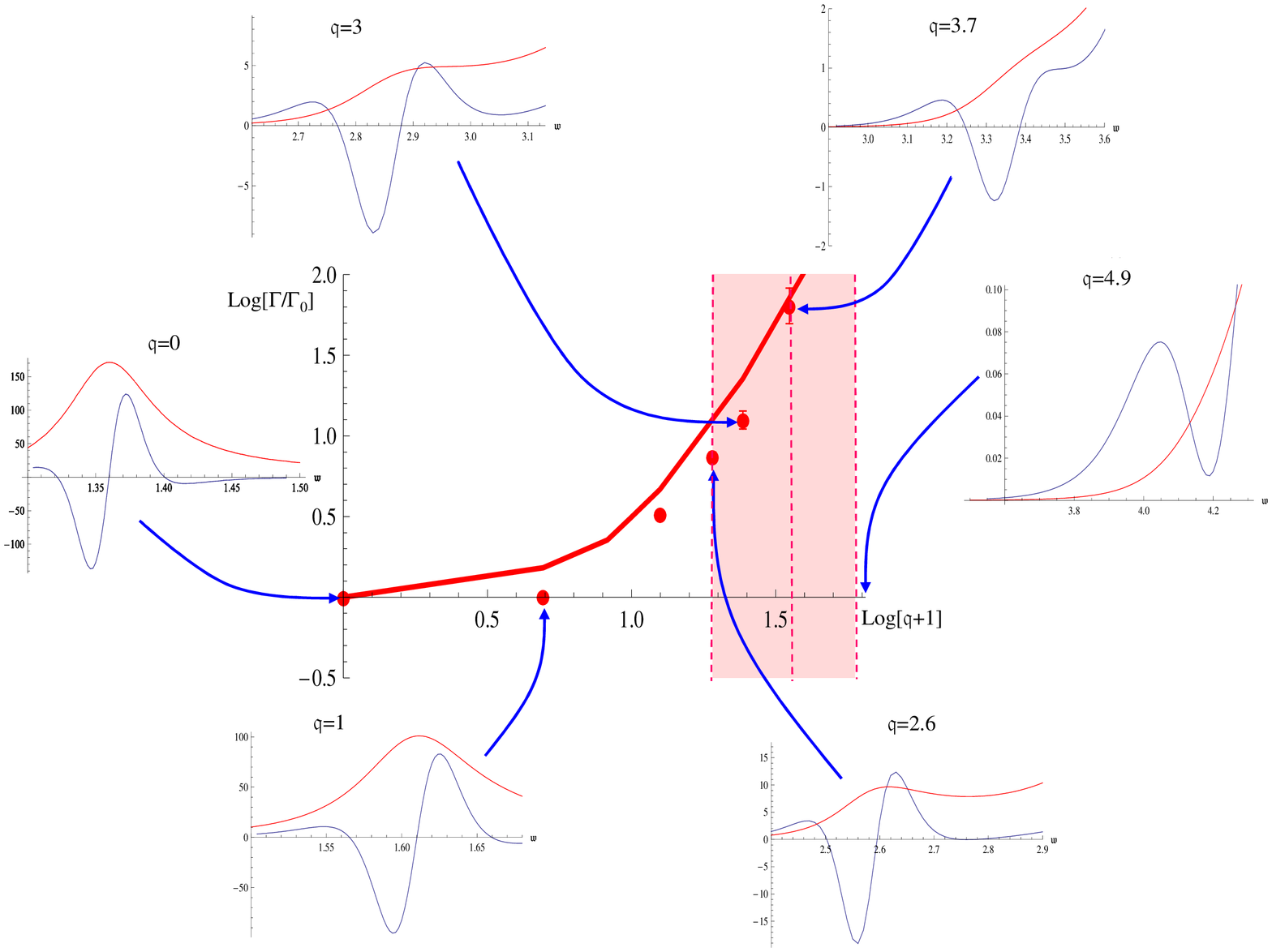}
\caption{The central plot shows the behaviour of $\Gamma$ with $\gq$
for case V as in fig.~\ref{Gammavsqfromplots}. For each point in
this plot, we also show the corresponding spectral function $\gR$
denoted by red and $\partial_\gw^3 \gR$ denoted by blue. This
typical example illustrates how the peaks lose their form with
increasing $\gq$.}\label{newone} }
\end{center}

Beyond the point where $\gw_0$ is lifted, there is no (simple) way
to use the spectral function to follow the corresponding pole in the
thermal correlator and certainly there is no sense in applying the
concept of a quasiparticle. One might use the lifting of the zero
$\gw_0$ as defining the value $\qm$ where the quasiparticles
disappear. These momenta for each of the four cases in
fig.~\ref{Gammavsqfromplots} coincides to the extreme right of the
corresponding shaded region. We return to these issues in the
discussion in section \ref{discuss}.

A clear trend that emerges from fig.~\ref{Gammavsqfromplots} above
is that the value of $\qm$ increases with increasing $\vm$. This is
already clear in comparing the results for cases IV, V and VI, all
of which start with $\Gamma_0\simeq0.04$. However, while $\vm$ is
certainly one parameter that changes between these three cases,
another parameter which distinguishes these three cases is
$\Gamma_0/\Omega_0$. The latter increases significantly between
these three cases while the limiting velocity is decreasing and one
might imagine that if the quasiparticle peaks begin by being less
well resolved, they should be washed out more quickly.  Hence we
also considered case VII for which $\vm=0.651$ and
$\Gamma_0/\Omega_0\simeq0.01$. So here the limiting velocity matches
that in case V (for which $\Gamma_0/\Omega_0\simeq0.03$) but  the
ratio $\Gamma_0/\Omega_0$ was tuned to match roughly that in case IV
(for which $\vm=0.995$). In fig.~\ref{Gammavsqfromplots}, we see
that the width for case VII grows with $\gq$ in close accord with
that for case V. We note though that our estimates of $\qm$ for case
VII were slightly larger than those of case V. In any event, these
results seem to indicate that $\vm$ is the dominant factor in
determining critical momentum where the quasiparticles disappear.

For comparative purposes, we also show in fig.~\ref{spectralgrid2}
the ratio of the widths of the first two peaks in case VI, as
calculated with \reef{next2}. As the plot shows, the width of the
second peak increases more slowly than the first. For instance,
$\Gamma_2(\gq)/\Gamma_1(\gq)\simeq 4.3$ at $\gq=0$ but only 1.3 at
$\gq=2.2$. This behaviour is in fact typical for all of the cases
which we studied here.
\FIGURE[ht]{
 \includegraphics[width=0.7 \textwidth]{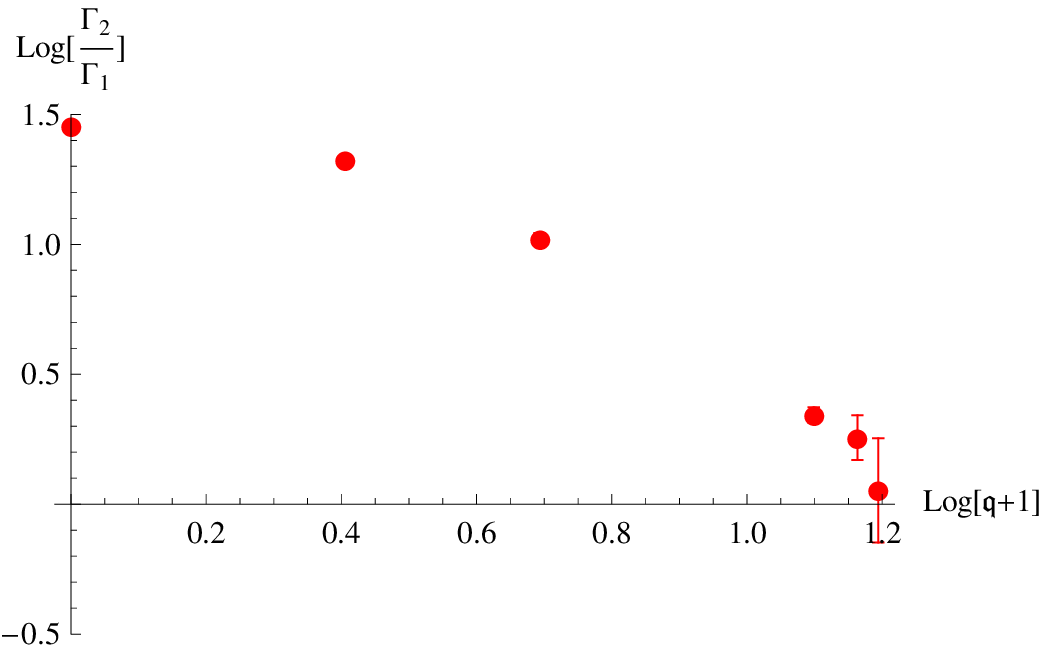}
\caption{The plot shows $\log (\Gamma_2/\Gamma_1)$ for the first two
peaks in case VI.  For these two peaks, $\Gamma_0$ takes the values
$0.042,0.184$ respectively. The first peak disappears at $\gq\approx
3$ while the second peak disappears at $\gq\approx 2.72$.}
\label{spectralgrid2}}
\vskip .5cm

\noindent\underline{\bf Residues, $A(\gq)$}: So far our discussion
of the quasiparticle peaks has focussed on their energy $\qo$ and
the width $\qg$, which correspond to the real and imaginary parts of
the position of a pole in the thermal correlator \reef{tard}. Here
we briefly turn to the residue $A$, as this also determines the size
of the peak corresponding to a given pole -- \eg for the canonical
shape \reef{peek}, the maximum value of spectral function is:
$\gR(\gw=\qo)=2A/\qg$. Assuming the peaks have this canonical shape,
the residue can also be determined in a number of different ways
from the spectral functions but we focus on the third derivative of
the spectral function following the above discussion for the form
given in eq.~\reef{peek3}. First, we can relate $A$ to the slope of
$\partial^3_\gw\gR$ at $\gw=\gw_0$:
 \be\labell{circ}
A_1=\left.{\qg^5\over48}\,\partial^4_\gw\gR\,\right|_{\gw=\gw_0}\ .
 \ee
This expression evaluated with $\bar\Gamma$ from eq.~\reef{next2} is
shown in fig.~\ref{Agrid} for the four cases in table \ref{table2}.
Alternatively, the residue $A$ can be determined by the value of the
extrema at $\gw=\gw_{m,\pm}$:
 \be\labell{squar}
A_{2,\pm}=\left. \pm {\qg^4\over48}\,
\frac{(1+\delta^2)^4}{\delta(1-\delta^2)}
\,\partial^3_\gw\gR\,\right|_{\gw=\gw_{m,\pm}}\ .
 \ee
Our first observation is that the results for all three of these
expressions seem to agree quite well. One might also note how
similar the results are for cases V and VII, which both have
$\vm=0.651$. The most significant effect apparent in
fig.~\ref{Agrid} is that $A$ decreases quite rapidly as $\gq$
increases. This effect is most dramatic in case IV where the results
for $A$ are fit well by an exponential:
$A\simeq140\,\exp[-0.077\,\gq]$. This rapid fall in $A$ with
increasing $\gq$ was the third effect which we identified as
contributing to the disappearance of the quasiparticle peaks in the
spectral function.
\begin{center}
\FIGURE{
\includegraphics[width=1 \textwidth]{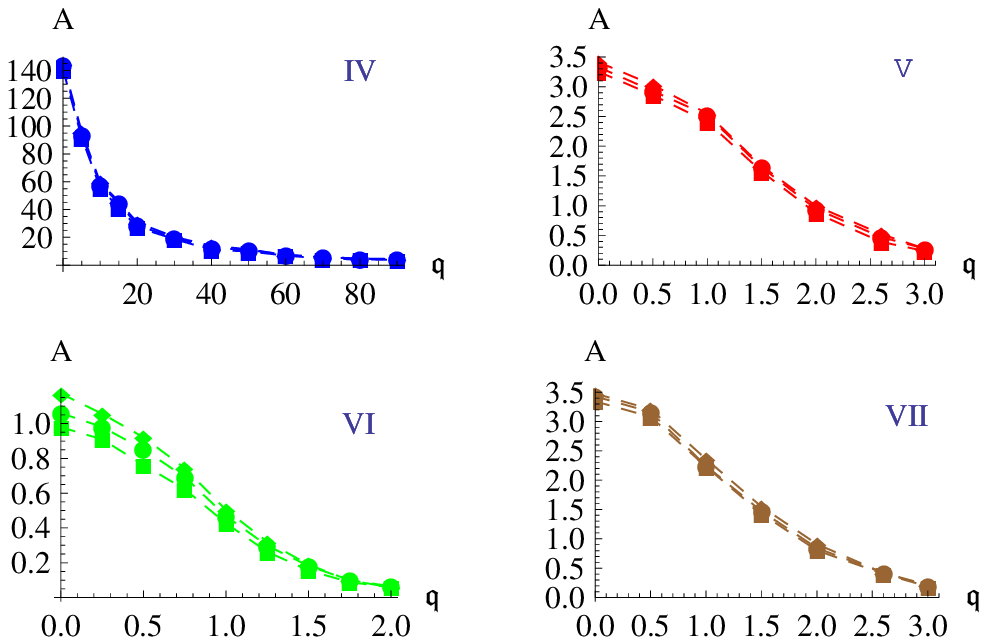}
\caption{The residue $A$ calculated for the first peak in the four
cases given in table \ref{table2}. The circles correspond to results
for $A_1$ in eq.~\reef{circ}, while the squares and diamonds
correspond to $A_{2,\pm}$ in \reef{squar},
respectively.}\label{Agrid} }
\end{center}

\section{Quasinormal modes}\label{quasi}

The basic features of the spectral functions are controlled by the
analytic structure of the corresponding retarded correlators in the
complex frequency plane. Holographically, these poles correspond to
quasinormal modes of excitations in the black hole geometry induced
on the D7-brane \cite{Son:2002sd,Birmingham:2001pj}. Investigating
the quasinormal spectrum for the black hole embeddings of interest
is a technically challenging problem which we intend to pursue
elsewhere \cite{prep}. However, as described in \cite{spectre}, some
qualitative information can be inferred from casting the relevant
radial equations of motion in the form of a Schr\"odinger equation.
In particular, one naively expects that the appearance of
quasiparticle peaks with $\Gamma\ll\Omega$ in the spectral function
would correspond to metastable states supported by a local minimum
in the effective potential of the Schr\"odinger equation.
Conversely, the absence of a local minima could be expected to yield
the non-existence of quasiparticles. In this section, we apply this
kind of qualitative analysis to the transverse vector and
pseudoscalar fluctuations of the D7-brane. These calculations allow
us to gain some qualitative insight into the quasinormal modes for
these fields -- see, \eg \cite{hoyos,nicknew}. One might be tempted
to apply a WKB analysis in this framework to obtain some
quantitative results as well, however, as we describe in appendix
\ref{wonk}, such an approach will not yield reliable results.

Let us begin with \reef{ETeom}, the equation of motion for the
transverse vector, and follow the approach described in appendix of
D of \cite{spectre}. Defining\footnote{Recall that $F$ and $G$ are
defined in \reef{core} and $\Delta$, in \reef{Delta}.} $H_0=(f^2
F\Delta)/(\tilde f^2 G)$ and $E_T=h\psi$, we find that choosing
$h=H_0^{1/4}/F^{1/2}$ recasts this equation into
 \be\label{schrod}
-\partial^2_{R_*} \psi+V\,\psi=\gw^2\,\psi\,,
 \ee
where $R_*\equiv\int_\rho^\infty d\tilde\rho/\sqrt{H_0(\tilde\rho)}$
and the effective potential $V$ is given by
 \ba
V&=&{f^2 \over \tilde f^2 }\Delta\left[-{1\over G\,
h}\partial_\rho(F\partial_\rho h)+\gq^2\right]\nonumber\\
&=& V_0+\gq^2\ V_1\,.\label{potter}
 \ea
Note that with the above definition for $R_*$, we have
$R_*\rightarrow \infty$ at the horizon (\ie $\rho\rightarrow1$) and
$R_*\rightarrow 0$ asymptotically (\ie $\rho\rightarrow\infty$). We
also note that the simple dependence of the Schr\"odinger problem on
the frequency and wave-number of the mesons. That is, $\gw^2$ plays
the role of the effective energy while $\gq^2$ appears as the
coefficient of a new term $V_1$ in the effective potential. Further
we observe that in the limit that $\rho\rightarrow\infty$,
$V_1\rightarrow 1$ and hence in this UV regime, these two
contributions can be recombined in the Lorentz invariant form
$\gw^2-\gq^2$.

We begin by illustrating the behaviour of $V_0$ in fig.~\ref{10x}.
We might compare this to the corresponding effective potential
discussed for the Minkowski embeddings in \cite{us-meson}. Both
there and here, the effective potential rises to infinity for large
$\rho$ which simply reflects the infinite gravitational potential of
the AdS geometry. In the Minkowski case, there is also an infinite
barrier at $\rho=R_0$ where the D7-brane embedding ends and so this
potential leads to a discrete spectrum of (stable) bound states. As
described in section \ref{hole}, the D7-branes of interest here with
$\dd$ have a narrow neck that extends down to the black hole horizon
at $\rho=1$. As a result, the infinite barrier at $\rho=R_0$ above,
is reduced to a finite barrier, as illustrated in fig.~\ref{10x}. As
a result, one's intuition should be that the low lying bound states
for the Minkowski problem remain essentially unchanged but they are
now only metastable because they will slowly tunnel out through the
potential barrier to the horizon. Further as shown in
fig.~\ref{10x}, as $\dd$ increases (with $m$ fixed), the height of
the potential barrier shrinks and so the decay rate of these
metastable states will increase. Of course, for highly excited
states with $\gw^2$ above the top of the barrier, the spectrum will
be completely changed and in particular, we do not expect to speak
in terms of metastable states.
\FIGURE{\begin{tabular}{cc}
\includegraphics[width=.45 \textwidth]{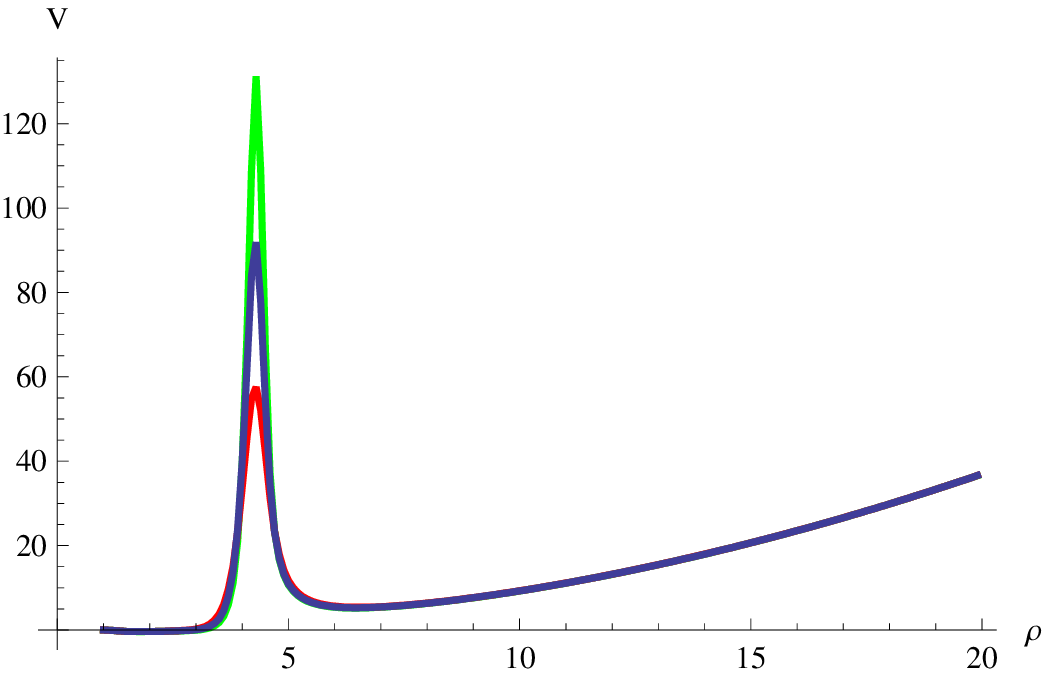}&
\includegraphics[width=.45 \textwidth]{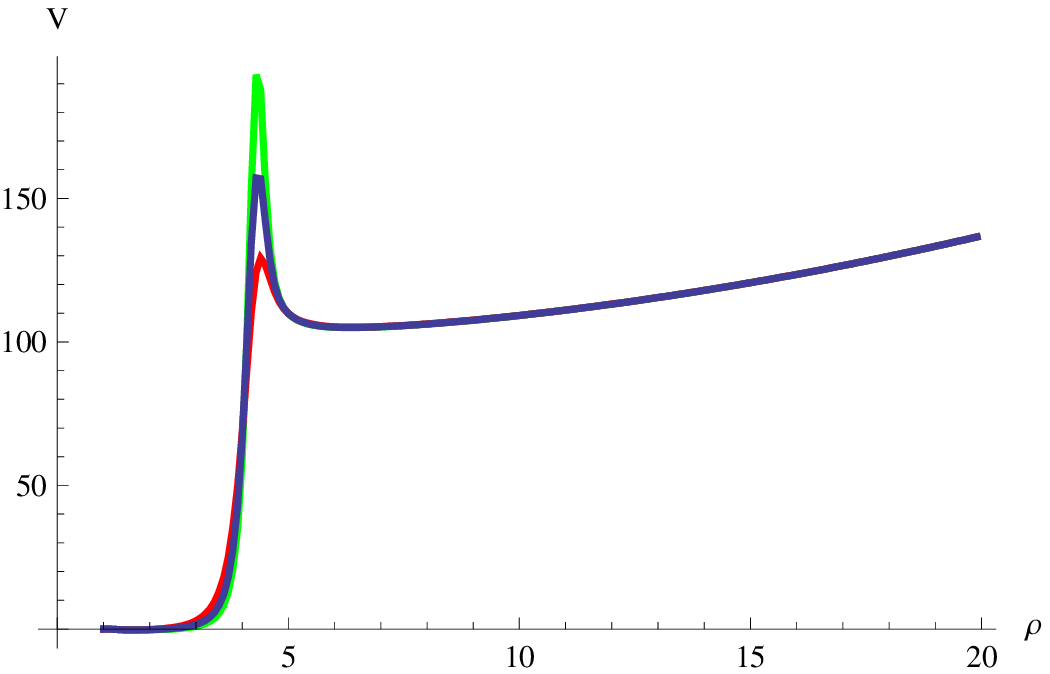}
\end{tabular}
\caption{The plot on the left is for $V_0$ vs $\rho$ for $\vm=0.995$
and $\tilde d=0.15$ (case IV; green), 0.25 (case I; blue), 0.5
(red). The potential barrier diminishes with increasing $\tilde d$.
The plot on the right shows the total potential $V$ with $\gq=10$
for the same range of parameters.}\label{10x}}

When the meson states also have a finite momentum, we must also
account for the contribution $\gq^2\,V_1$ to the effective
potential. Given the expression in \reef{potter}, one finds that
this term raises the effective potential by a finite amount are
larger values of $\rho$. As illustrated on the right in
fig.~\ref{10x}, while both the local maximum and minimum in the
potential are raised, the primary effect of increasing $\gq$ is to
raise the minimum relative to the maximum, \ie the barrier shielding
the potential well from the horizon is reduced. This effect is also
illustrated in fig.~\ref{Vk} where we have plotted $V-\gq^2$ with
changing $\gq$ -- subtracting $\gq^2$ ensures that the asymptotic
form of the resulting curves is fixed, as can be inferred from the
discussion below \reef{potter}.
\FIGURE{
 \includegraphics[width=0.7 \textwidth]{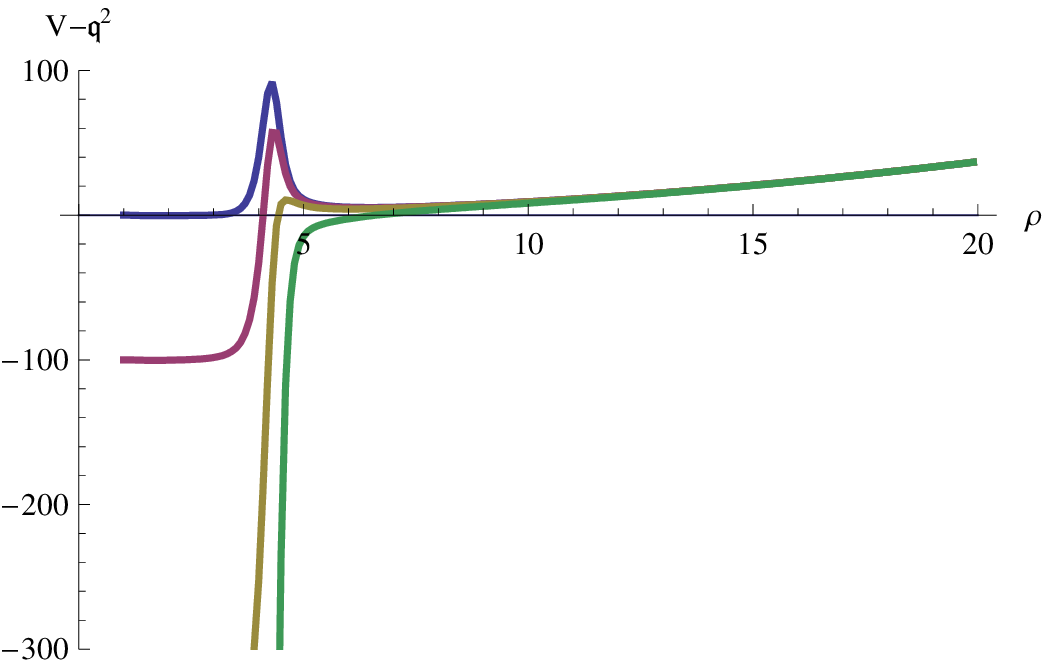}
\caption{$V-\gq^2$ vs $\rho$ for case I (\ie $\tilde d=0.25$ and
$\chi_0=0.99$). The curves correspond to $\gq=0$ (blue), $10$
(pink), $20$ (yellow) and $50$ (green). As $\gq$ increases the
barrier decreases and eventually disappears. } \label{Vk}}

To summarize, recasting the equation of motion in the Schr\"odinger
form \reef{schrod} gives an intuitive picture which yields the
following key observations: For small $\gq$, the effective potential
has a large barrier, which is responsible for the metastability of
the quasiparticles. However, this barrier shrinks with increasing
$\gq$ which then explains why the decay rates of the metastable
states should increase as $\gq$ increases. Further this barrier
actually disappears above some momentum which again points to a
critical momentum $\qm$ beyond which the quasiparticles disappear.

We can use this analysis to establish a quantitative criterion for
this maximum momentum $\qm$. Since our intuition is that metastable
states exist because of the potential barrier appearing in $V$, we
propose to estimate $\qm$ as the value of the momentum at which the
barrier disappears and is reduced to a point of inflexion. We have
shown these estimates in fig.~\ref{Gammavsqfromplots} for the three
cases described in table \reef{table2}. It seems that for cases V,
VI and VII (with the smaller values of $\vm$), these estimates are
well within the region where the peaks are losing their canonical
shape and in fact, are quite close to the point where the zero at
$\gw_0$ is lifted. The latter was also suggested as a measure of
$\qm$ in the previous section. For case IV with $\vm=.995$, the
result here gives a smaller value of $\gq$ outside of the region
where peak is losing its shape and so present calculation seems to
underestimate $\qm$. However, it still seems to coincide with the
onset of the rapid rise in $\Gamma$. It seems somewhat surprising
though that the widths could still be relatively small after the
barrier in the effective Schr\"odinger potential had disappeared.

In any event, the advantage of the present approach to estimating
$\qm$ is that it is computationally simple and relatively
inexpensive. For example, in fig.~\ref{11x}, we used this approach
to determine the behaviour of $\qm$ as a function of $T/\bar M$. As
$T/\bar M$ essentially fixes $\vm$, these results are again
displaying the strong correlation between $\qm$ and the limiting
speed discussed in the previous section. In fig.~\ref{11x}, the
continuous curves were constructed by fitting the data points with
the form $\exp[-8x]/x^2$. Hence we find in the zero temperature
limit, $\qm\propto(\bar M/T)^2\rightarrow \infty$. Furthermore, in
the limit that $\dd\rightarrow 0$ (and hence $\Gamma_0\rightarrow
0$), $\qm\rightarrow \infty$.
\FIGURE{
\includegraphics[width=0.85 \textwidth]{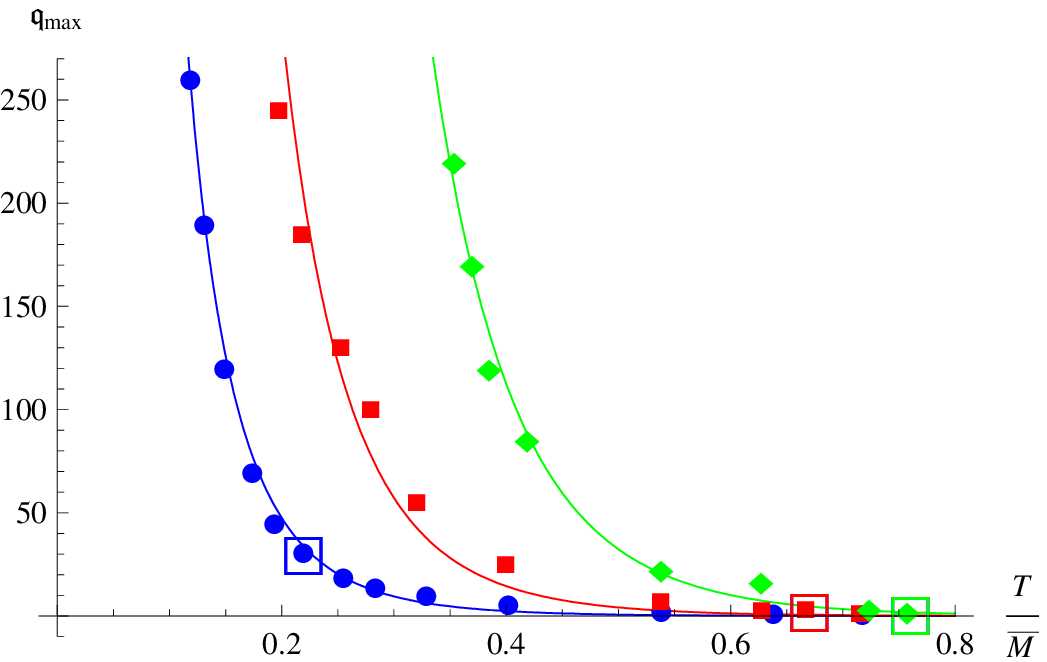}
\caption{This plot shows $\qm$ vs $T/\bar M$ for $E_T$. The
 blue, red and green points and curves correspond to  $\tilde
d=0.25$, 0.0005 and 0.0001, respectively. Further the boxes surround
the points corresponding to cases I, II and III in table
\reef{table}. The continuous curves come from fitting the data
points with the form $\exp[-8x]/x^2$.}\label{11x} }

We close this section with a few technical comments. One may have
imagined that the decay widths would be calculable in the
corresponding Schr\"odinger problem using a WKB approximation. Such
a calculation is outlined in appendix \ref{wonk}. However, as also
explained there, this approach does not generally yield reliable
results. The essential point is that for the WKB approximation to be
valid, the change of the momentum over a wavelength must be small
compared to the momentum itself. Unfortunately, in the examples
considered here, this condition in not met within the barrier making
the WKB analysis unreliable.

Finally we comment briefly on the effect on the widths from adding
angular momentum on the internal $S^3$. That is, all of mesons
considered up to this point have been singlets under the internal
$SU(2) \times SU(2)$ global symmetry of the gauge theory -- this
symmetry is dual to rotations on the D7-brane's internal $S^3$.
However, these states only correspond to the lowest dimension
operators in an infinite family of vector operators transforming in
the ($\ell/2$, $\ell/2$) representation of the internal symmetry
\cite{us-meson}. So we wish now to consider states with nonvanishing
$\ell$. In fact, we will turn to the pseudoscalar mesons at this
point since the analysis in this case is somewhat simpler. For
non-zero $\ell$ and $\gq$, $A_\rho$ cannot generally be set to zero
\cite{spectre} which complicates the analysis for the vector.
However, adding angular momentum to the pseudoscalar case is
straightforward and so we consider this case here. The equation of
motion for the pseudoscalars, \ie fluctuations in $\phi$, can be
written as
 \be
\partial_\rho(F\partial_\rho\,{\cal P})+G(\gw^2{\tilde f^2\over \Delta f^2}
-\gq^2){\cal P}+\ell(\ell+2) H\, {\cal P}=0\,,
 \ee
where the functions $F,G,H$ are defined as
\begin{eqnarray}\label{pscalar}
H_0&=& {\rho^4 f^2\over 8\tilde f} {1-\chi^2\over
1-\chi^2+\rho^2\chi'^2}\,,\quad F={\rho^5 f\tilde f
\chi^2(1-\chi^2)^2\over \sqrt{\Delta}
\sqrt{1-\chi^2+\rho^2\chi'^2}}\,,\\
H_2&=& \Delta {f^2\over \tilde f^2}\,,\quad H_3= {\rho^2 f^2 \Delta
\over 8 \tilde f (1-\chi^2)}\,,\quad G= {F H_2\over H_0}\,,\quad
H={F H_3\over H_0}\,.
\end{eqnarray}
It can be shown that the effective Schr\"odinger potential in this
case can be written as \cite{spectre}
 \ba\label{potter2}
V&=&V_0+\gq^2\,V_1+\ell(\ell+2)\,V_2\\
&=&-{H_0\over h F}\partial_\rho (F\partial_\rho h)+\gq^2\,
H_2+\ell(\ell+2)\, H_3\,,
 \ea
where $h=H_0^{1/4}/F^{1/2}$. Some examples of the effective
potential are shown in fig.~\ref{12x}. The key feature which one
observes from this plot is that introducing $\ell\ne0$ increases the
potential barrier. This is not surprising since the D7-brane
embedding has a narrow neck near the horizon, \ie the $S^3$ is
small, and so this increase reflects an angular momentum barrier for
these modes. We also note that the plots of the potential for the
vector meson $E_T$ for $\ell=0$ are almost indistinguishable from
those for the pseudoscalar even for non-zero $\gq$. In fact the two
potentials differ by less than $1\%$ although their precise
functional forms are not identical.
\FIGURE{
\includegraphics[width=0.75 \textwidth]{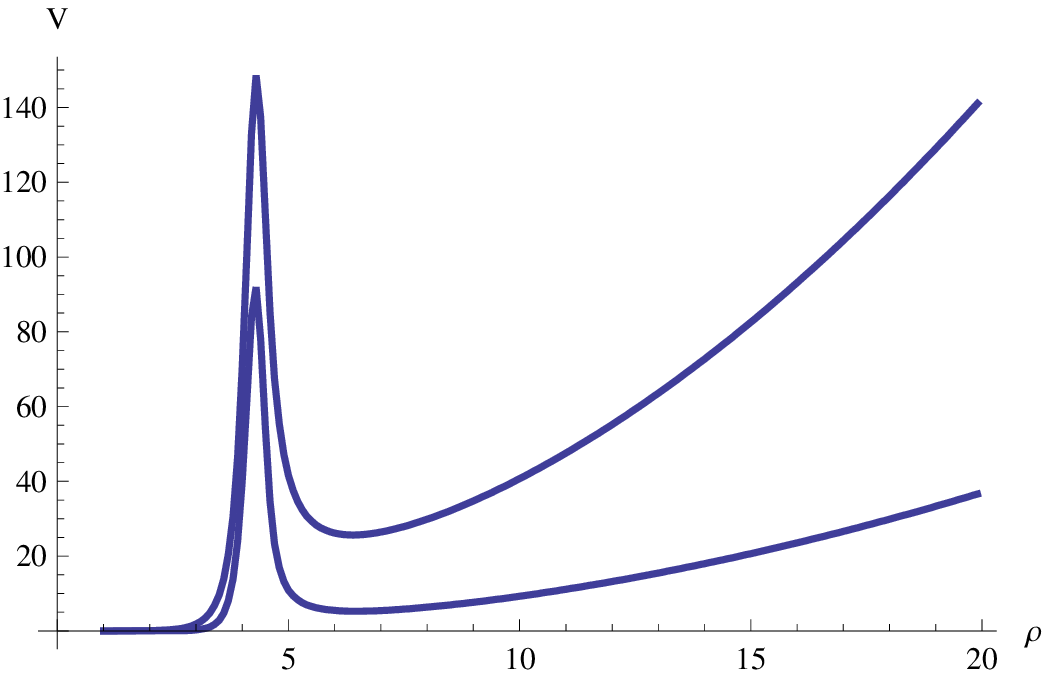}
\caption{This plot is for the effective Schr\"odinger potential for
the pseudoscalar field with $\ell=0,1$ and $\gq=0$, $d=0.25$,
$\chi_0=0.99$. The higher potential is for $\ell=1$.}\label{12x} }

\section{Beyond $\qm$} \label{beyond}

As we saw in section \ref{spectral}, the analysis of the spectral
functions can only give limited information about the dispersion
relations. In particular, when the quasiparticles become too
unstable with increasing $\gq$, they no longer contribute
characteristic features to the spectral function which would allow
us to infer $\Omega(\gq)$ and $\Gamma(\gq)$. Of course, this simply
indicates that the corresponding pole in the thermal correlator
\reef{tard} has moved down in the complex plane, too far from the
real axis to strongly influence the spectral function. The behaviour
of these poles can be followed to higher momenta in the present
holographic framework with a direct analysis of the corresponding
quasinormal modes \cite{quasi,andrei1,andrei2}. While we intend to
pursue this problem elsewhere \cite{prep}, here we will attempt to
provide some qualitative insight into the behaviour of the poles at
higher momentum and also for the higher quasinormal modes, using the
intuitive picture of the effective Schr\"odinger problem given in
the previous section.

Our intuitive picture will also be guided by a comparison with the
results for the $R$ currents in ${\cal N}=4$ SYM at finite
temperature. That is, we start by considering the behaviour of the
transverse vector modes of a Maxwell field in $AdS_5$. A closely
related investigation of the longitudinal vector modes and a
massless minimally coupled scalar appeared in \cite{andrei1} and
\cite{andrei2}, respectively.\footnote{There is an erroneous claim
in \cite{andrei1} that the results for the transverse vector modes
should be precisely the same as for that for the massless minimally
coupled scalar field given in \cite{andrei2}.} The relevant equation
of motion for the transverse vector modes $E_T$ in $AdS_5$ can be
found in \cite{spectre, quasi} and can be written as
 \be\label{sugreom}
\partial_\rho\left(\rho^3f\partial_\rho E_T\right)
+{8\over\rho}\,{f\over \tilde f}\left(\gw^2{\tilde f^2\over
f^2}-\gq^2\right) E_T=0\,,
 \ee
where $f$ and $\tilde f$ are defined above in \reef{ffff}. This
equation of motion is equivalent to that of the analogous modes of
the D7-brane gauge field \reef{ETeom} with $\chi=0$ and $\dd=0$.
Hence using the results of section \ref{quasi}, eq.~\reef{sugreom}
is easily recast into the Schr\"odinger form
 \be\label{schrod2}
-\partial^2_{R_*} \psi+V(\rho)\,\psi=\gw^2 \psi\,,
 \ee
where $R_*=\int_{\rho}^\infty d\tilde\rho/\sqrt{H_0(\tilde\rho)}$
with $H_0(\rho)=\rho^4f^2/8\tilde f$. The wave-function has been
defined as $\psi\equiv E_T/h$ with $h=(8\rho^2\tilde f)^{-1/4}$ and
the effective potential $V(\rho)$ is given by
 \be \label {sugrapot}
V(\rho)=V_0+\gq^2\ V_1={f^2\over \tilde f^2}
\left[{3\over32}\,\rho^2\tilde f+{5\over8}\,{1\over\rho^2\tilde
f}+\gq^2 \,\right]\,.
 \ee
Again the effective potential in eq.~\reef{potter} reduces to that
above upon setting both $\chi$ and $\dd$ to zero. Fig. \ref{sugraV}
presents various plots of the effective potential versus the
Schr\"odinger coordinate $R_*$ for various values of $\gq$. In fig.
\ref{comp}, we also compare this potential \reef{sugrapot} for the
supergravity vector modes with that for the analogous D7-brane modes
for different values of the quark mass and $\dd$.
\FIGURE[h]{
\includegraphics[width=0.8 \textwidth]{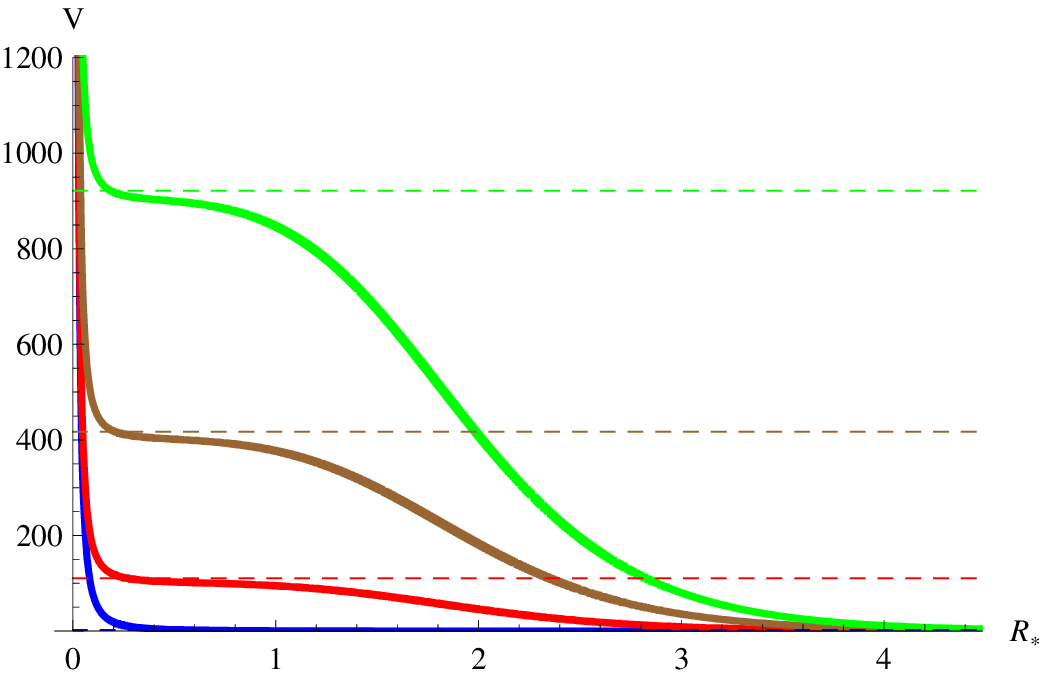}
\caption{The plot shows $V-\gq^2$ vs $R_*$. Blue, red, brown and
green are for $\gq=0,10,20,30$ respectively. The dashed lines
indicate the lowest eigenfrequency obtained from the spectral
function at these $\gq$'s.} \label{sugraV}}

As noted above, \cite{andrei1,andrei2} presented an analysis of
quasinormal modes for similar supergravity fields in $AdS_5$. Here
in our preliminary analysis of the transverse vector, we use only
the spectral function techniques described in section \ref{spectral}
for the lowest lying modes. The analogue of eq.~\reef{FTeom} is
given here by
\be
\partial_\rho \gF +{\gw^2\over\rho^3\, f}
\gF^2+{8\over\rho}\,{f\over \tilde f}\left( {\tilde f^2\over
f^2}-{\gq^2\over\gw^2} \right)=0\,, \ee
where $\gF$ is defined in \reef{speck} with $F=\rho^3f$. The
boundary condition for regularity at the horizon becomes
$\gF(1)=-{4i\over \gw}$ and the spectral function is given by the
asymptotic limit in \reef{speck2}.

\FIGURE[ht]{
\begin{tabular}{cc}
\includegraphics[width=0.45 \textwidth]{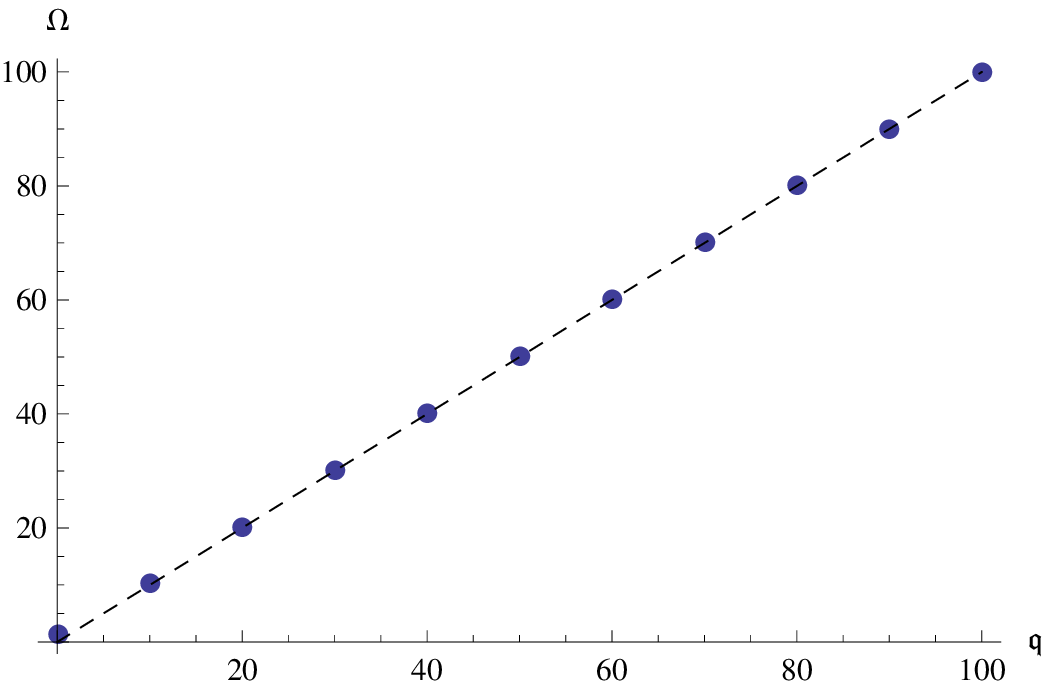}
&
\includegraphics[width=0.45 \textwidth]{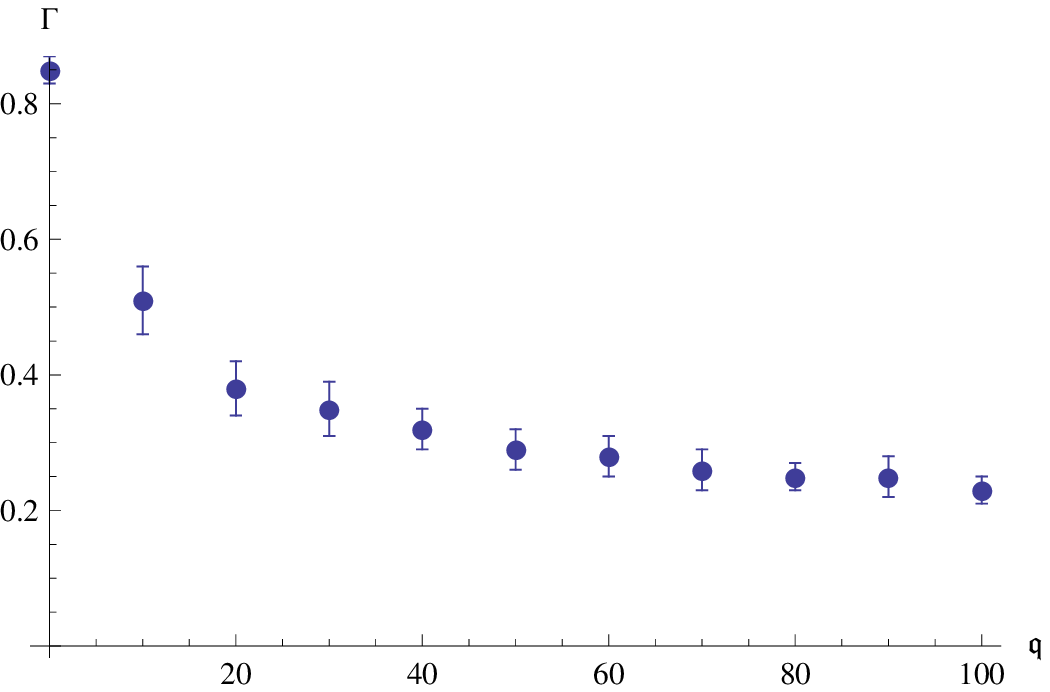}\end{tabular}
\caption{The energy $\Omega(\gq)$ and width $\Gamma(\gq)$ for the
lowest lying mode for the supergravity gauge field. The black dashed
line on the right corresponds to $\Omega=\gq$.} \label{sugraomega}}
Next we can repeat the analysis of section \ref{disperse} to extract
$\Omega(\gq)$ and $\Gamma(\gq)$ from the spectral function -- in
particular, our results were derived using the structure of
$\partial^3_\gw\gR$. The behaviour of the lowest lying mode are
plotted in fig. \ref{sugraomega}. The most striking features
revealed there are a) the asymptotic velocity is precisely one and b)
the width decreases with increasing $\gq$. Hence there is a
remarkable contrast between this behaviour for the supergravity
vector and that described for the D7-brane vector in section
\ref{disperse}.

We should note that at $\gq=0$ our approach here yields
$\left(\Omega(0),\Gamma(0)\right) =(1.51,.85)$ while the precise
analytic results are known to be
$\left(\Omega_n(0),\Gamma_n(0)\right) =(n,n)$ \cite{spectre}. Hence
we have an indication of the potential errors in our spectral
function results. We should note that an analytic expression for
spectral function at $\gq=0$ is also known \cite{spectre} and our
numerical calculation of spectral function agrees well with this
result. However, it is not surprising that our technique for
extracting the energy and width from $\gR$ is giving imprecise
results here since we have $\Omega_n(0)=\Gamma_n(0)$ and further
$\Delta\Omega\simeq\Gamma_n$ for the first few poles. We remain
confident that the qualitative behaviour appearing in fig.
\ref{sugraomega} is correct. In particular, precisely the same
behaviour was first found for the supergravity modes studied in
\cite{andrei1,andrei2}, \eg see figures 3 and 4 in \cite{andrei2}.

We would like to understand these qualitative features from the
structure of the effective potential in the Schr\"odinger equation
\reef{schrod2}. As shown in fig. \ref{sugraV}, one of the
interesting features of $V$ is that it develops a relatively flat
plateau  in the regime $0<R_*<1$, which corresponds to large values
of $\rho$. Further, calculating the real part of the effective
energy $\Omega_S={\rm Re}(\gw^2)=\Omega^2-\Gamma^2$ -- see appendix
\ref{wonk} -- for the lowest lying level, we find that this energy
is just above the potential energy at the plateau. If we consider a
WKB calculation of the corresponding wave-function, our intuition is
confirmed about both of the qualitative features above. In
particular, in this approximation, the amplitude of the
wave-function would be given by \cite{qbook}
 \be\labell{ample}
|\psi|^2\simeq {1\over P(R_*)}={1\over\sqrt{\Omega_S-V(R_*)}}
 \ee
where $P(R_*)$ is the classical momentum of the particle (with
energy $\Omega_S$) evaluated at the position $R_*$. Hence the
amplitude of the wave-function is largest on the plateau and so it
seems natural that the wave-function has its largest support in this
region $0<R_*<1$. For large $\rho$, we have $R_*\sim2\sqrt{2}/\rho$
and so the latter range corresponds to large $\rho$. Hence the
radial profile of the corresponding excitation has its support
primarily in the region where the geometry is very close to that of
$AdS_5$ and the redshift effects are minimal. Hence it is not
surprising that the asymptotic velocity of these excitations is one,
\ie the speed of light.

Quantitatively we may also observe from fig. \ref{sugraV} that
$V_{plateau}\sim \gq^2$ and so with a small width, we naturally have
$\Omega_S \simeq \Omega^2\simeq\gq^2$. Finally while $V_{plateau}$
and $\Omega_S$ are rapidly rising with $\gq$, the value of the
potential remains fixed at zero at the horizon which corresponds to
$R_*\rightarrow\infty$. Hence eq.~\reef{ample} indicates that the
relative amplitude of the wave-function at the horizon is falling
$|\psi|^2\sim 1/\gq$. Hence it seems natural that the flux absorbed
by the horizon should also be decreasing and hence the width of the
state would be decreasing.

\FIGURE[h]{\begin{tabular}{cc}
\includegraphics[width=0.45 \textwidth]{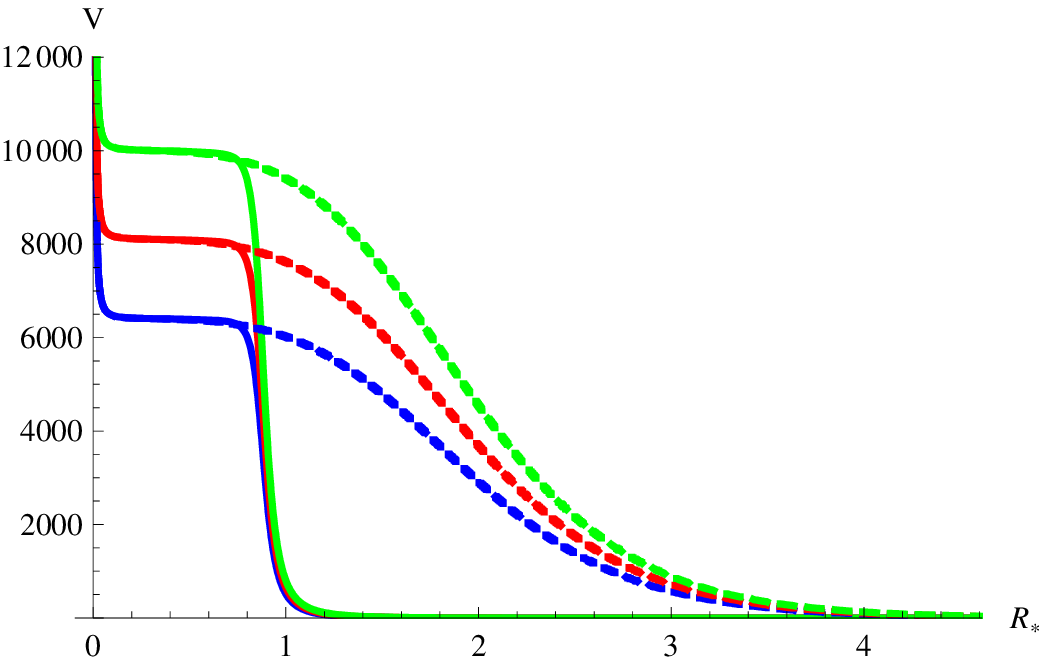} &
\includegraphics[width=0.45 \textwidth]{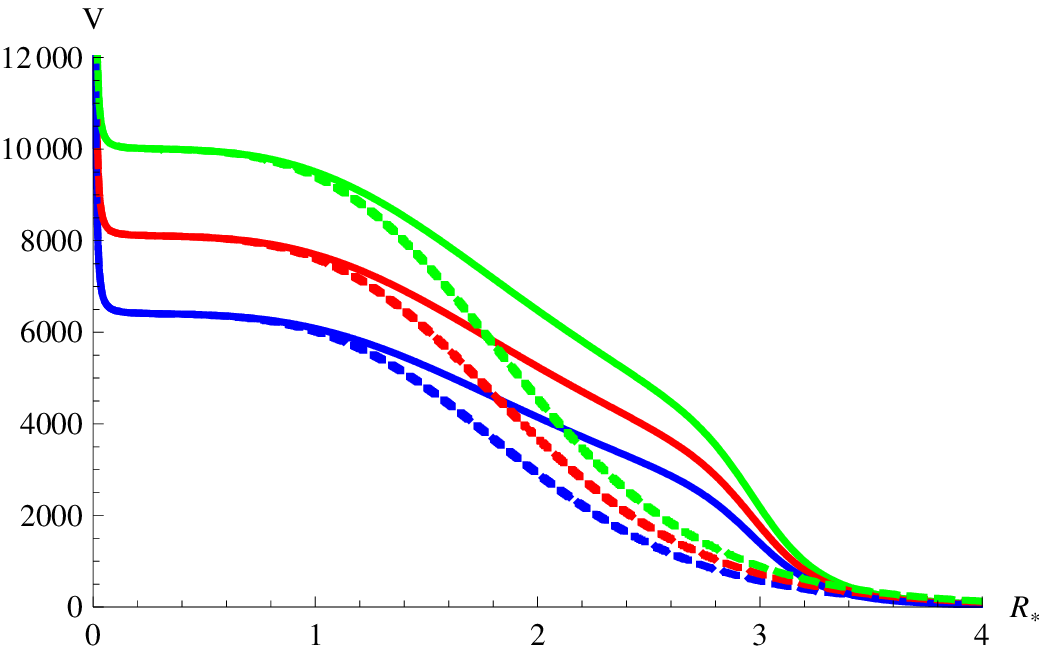} \end{tabular}
\caption{The solid lines correspond to the effective potential in
case I (left) and V (right) for $\gq=$ 80 (blue), 90 (red), 100
(green). The dashed lines correspond to the analogous supergravity
potential in (\ref{sugrapot}).}\label{comp} }
In principle, one could verify this intuitive picture with more
detailed calculations to produce a better understanding of the
precise normalization of the wave-function.\footnote{In the case of
near-critical embeddings, a flat plateau in the potential produced
similar effects in the quasinormal spectrum at $\gq=0$
\cite{gong}.} However, we now wish to apply our newly gained
intuition to infer the behaviour of the meson dispersion relations
at very large momentum. In fig. \ref{comp}, we compare the
supergravity potential \reef{sugrapot} to that for the analogous
vector modes on the D7-brane for two of the cases (I and V) studied
in section \ref{spectral}. In both cases, we are in a high-momentum
regime where $\gq\gg\qm$ and so there is no evidence of the
structure (\ie a potential barrier and well in front of the horizon)
discussed in section \ref{quasi}. While the structure of the
potentials in fig. \ref{comp} differs for $R_*>1$ (\ie deep in the
bulk of the $AdS_5$ black hole), they are essentially identical for
$R_*<1$ (\ie in the asymptotic region). In particular, the effective
potential for the mesons also develops a plateau with
$V_{plateau}\sim \gq^2$ in this region. Hence it is natural to
assume that at very high momentum, the poles characterising the
thermal correlator for the meson operators will display a behaviour
very similar to that found in the supergravity analysis. That is,
$\Omega(\gq)$ should exhibit an asymptotic velocity of one at very
large $\gq$ and the widths $\Gamma(\gq)$ should decrease as $\gq$
increases to very large values.

Hence we are led to conjecture that the full dispersion relations
found by studying the quasinormal modes dual to the meson operators
might take a form as illustrated in fig. \ref{possibility}. In
particular, the behaviour of the low-lying modes found for $\gq<\qm$
in section \ref{spectral} should not be indicative of the overall
structure. If we consider higher modes, \ie modes which do not lie
in the potential well of section \ref{quasi}, we expect that these modes will not exhibit a linear dispersion
  relation with slope $\vm$ in the $\gq<\qm$ regime as the low-lying modes
do.
  Rather their dispersion relations will directly approach a linear
  behaviour in the $\gq\gg \qm$ regime with an asymptotic velocity of
  the speed of light.
 Similarly their widths are likely to be monotonically
decreasing as found for the supergravity modes in
\cite{andrei1,andrei2}. We have found some evidence of this
qualitative behaviour for the higher modes from the spectral
functions but we do not believe our results are quantitatively
precise. There remains the interesting question of what should be
the high-momentum behaviour of the poles corresponding to the
low-lying modes, \ie those which exhibit $\vm<1$ for $\gq<\qm$. The
behaviour suggested in fig. \ref{possibility} is that beyond
$\qm$, $\Omega(\gq)$ should also reach $v_{\rm asym}=1$ for $\gq\gg\qm$.
However, we can only speculate on precisely how this should arise.
One possibility illustrated in fig.~\ref{possibility} is that
$\Omega(\gq)$ should rise up and asymptote to $\Omega=\gq$, which
would be similar to the behaviour exhibited by the sound mode in
\cite{quasi} -- see their fig. 5. This would require passing
through a regime where $\partial_\gq\Omega>1$, \ie the group
velocity appears to be superluminal, but of course, this need not
correspond to a violation of causality -- for example, see
\cite{superfast}. Another alternative also shown in
fig.~\ref{possibility} is that these modes approach
$\partial_\gq\Omega|_{lim}=1$ asymptotically without passing through
a regime with $\partial_\gq\Omega>1$.
\FIGURE[h]{
\includegraphics[width=0.85 \textwidth]{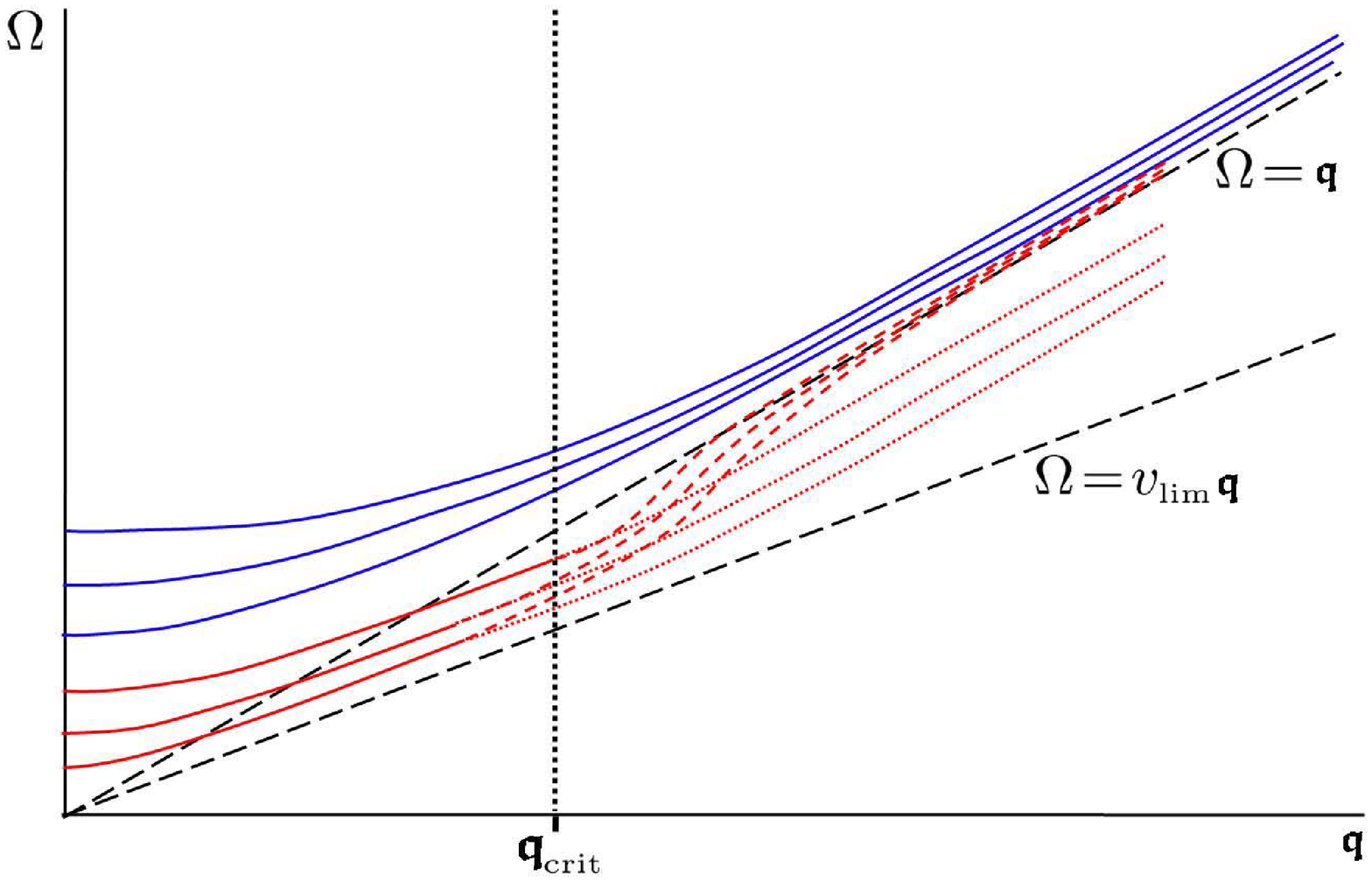}
\caption{This plot shows possible behaviours of the dispersion
relations beyond $\qm$. The lowest lying modes (red) exhibit $\vm<1$
below $\qm$ but approach $\partial_\gq\Omega=1$ in the regime
$\gq>\qm$. The latter limiting behaviour may (dashed curves) or may
not (dotted curves) exhibit `superluminal' speeds, i.e., $\partial
\Omega/\partial \gq >1$. The higher modes (blue) directly approach
the limit $\partial_\gq\Omega=1$ in the regime $\gq>\qm$.
\label{possibility}}}

We add one further observation on the spectrum of the quasinormal
eigenfrequencies. The poles corresponding to the quasiparticles
discussed in section \ref{quasi} corresponds to modes with support
primarily behind the barrier in the effective potential
\reef{potter}. Further the discussion here also refers to modes with
support primarily in the asymptotic region of the D7-brane. That is,
the support of all of these modes is mainly above the narrow throat
in D7-brane geometry. However, at least in the regime $\gq<\qm$,
there should be additional poles corresponding to quasinormal modes
whose support is predominantly in front of the potential barrier,
\ie in the narrow throat. These would be interpreted excitations of
the effective string gas that is modeled by the throat
\cite{finitemu}. These would be expected to have $\Gamma\sim\Omega$
and so would not lend themselves to a quasiparticle interpretation
\cite{seat}. It would also be interesting to understand the
behaviour of these quasinormal modes and their interplay with the
modes with support above the throat as we move into the regime
$\gq>\qm$. Certainly, as shown in fig. \ref{comp}, the structure
of the effective potential differs from that in the supergravity
problem for large $R_*$ (small $\rho$), hence it could be that it
still supports an additional set of modes in this regime with
distinct physical characteristics.

\section{Discussion}\label{discuss}

In this paper, we have used holographic techniques to calculate the
spectral functions and to study the dispersion relations of meson
quasiparticles moving through a thermal plasma of a strongly coupled
${\cal N}=2$ super-Yang-Mills theory. The quasiparticle peaks in the
spectral functions arise from poles in the corresponding thermal
correlator \reef{tard} at $\Omega-i\Gamma$ in the lower half of the
complex frequency plane. Considering the energy $\Omega$ and the
width $\Gamma$ as functions of the spatial momentum $\gq$, the broad
picture which emerged was that there were two distinct regimes
distinguished by a critical momentum $\qm$. In the low-momentum
regime $\gq<\qm$, the spectral function exhibited clear
quasiparticle peaks. By following their position and shape with
growing $\gq$, we were able to estimate the the dispersion
relations, $\Omega(\gq)$ and $\Gamma(\gq)$, for the low-lying
resonances. In the high-momentum regime $\gq>\qm$, the poles have
moved too far into the complex plane and no quasiparticle peaks are
evident in the spectral functions. However, we examined the
effective Schr\"odinger equation governing the dual quasinormal
modes and provided some qualitative insight into the dispersion
relations in this regime.

One of our key results for the $\gq<\qm$ regime is that the
quasiparticles approach the same limiting velocity found for the
case of stable mesons \cite{long,ejaz}. This result emphasizes that
this limiting velocity will be a universal feature in any
holographic model. While the precise form of the limit may depend on
the particular model, it arises from gravitational redshifting in
the background geometry, as indicated by \reef{speed}. Hence it will
apply for any gauge theory excitations that have a dual description
in terms of {\it radially localized} modes in the dual geometry. So,
for example, this robust feature would also apply in the adjoint
sector to `glueball' excitations that are dual to a wavepacket of
supergravity modes localized in the radial direction.

Of course, not all excitations of interest in a holographic model
need not be radially localized. For example, massive quarks in the
${\cal N}=2$ super-Yang-Mills theory (with $\nq=0$) are represented
by extended strings stretching down from the D7-brane to the horizon
in the dual gravitational description. Of course, the same
gravitational redshift is observed to have interesting physical
effects when these quarks are in motion -- \eg see \cite{flurry}.
Further, using holographic Wilson lines, a similar effect was
observed to lead to the dissociation of a heavy quark bound state
moving at a finite velocity through a strongly coupled ${\cal N}=4$
plasma \cite{wind}.

The second interesting observation for the low-lying mesons in the
$\gq<\qm$ regime is that their widths show a dramatic increase as
the momentum approaches $\qm$. In our holographic gravity model, the
rise in $\Gamma$ is easy to understand. As observed for the
Minkowski embeddings \cite{long}, with finite $\gq$, the excitations
on the probe D7-brane feel an extra potential which pushes the
support of the radial profile of the mesons down to minimum radius
on the brane. Intuitively, this can be understood as when the
momentum is introduced, the brane fields carry extra local energy
density and so feel a stronger gravitational force pulling them
towards the horizon. As described in section \ref{hole}, we have
destabilized the mesons by introducing a narrow throat which extends
down to the horizon from the point where the brane would have
otherwise closed off. Hence the additional potential due to the
momentum naturally pushes the meson fields down this throat making
these states decay more quickly. This intuitive picture can be made
more quantitative with the Schr\"odinger framework introduced in
section \ref{quasi}. The latter approach also made clear that the
potential barrier, which maintained the metastability of the
quasiparticles, vanishes above some momentum. The disappearance of
the barrier in the effective potential provides a precise
mathematical criterion with which to define $\qm$. While this is a
natural definition, we should add that the distinguishing momentum
$\qm$ remains a rather qualitative concept and another definition
was also considered in section \ref{spectral}.

In the high-momentum regime, our qualitative investigation of the
effective Schr\"odinger equation suggests that the asymptotic velocity is $v_{\rm asym}=1$ for
$\gq\gg\qm$. Here the modes should be radially localized but evade
having a limiting velocity less than the speed of light because they
are not localized near the black hole horizon. Rather, our
suggestion is that as $\gq$ increases to very large values, the
support of quasinormal modes becomes increasingly focussed at very
large radius, \ie towards the AdS boundary, where the gravitational
redshift becomes vanishingly small. This conjecture is largely based
on making an analogy with similar supergravity fields for which some
explicit results for the quasinormal frequencies are known
\cite{andrei1,andrei2}. The latter results further suggest that in
this high momentum regime the widths should decrease with increasing
momentum. To clarify the details of the behaviour in this
high-momentum regime, one would have to examine the quasinormal
modes directly \cite{prep}. It would also be interesting to better
understand the physical differences between the present case and
that with $\nq=0$ in which the mesons are display $\vm<1$ for
arbitrarily large $\gq$.

Still given the present understanding, an interesting physical
picture has emerged. Namely, in the low-momentum regime, the
quasiparticles are strongly coupled to the deconfined plasma of the
adjoint fields. The gravitational redshift leading to $\vm<1$ is the
geometric description of this strong coupling. An interesting
problem would be to map the `glue' cloud associated with these
moving mesons, \eg along the lines of \cite{news}. We expect that
the energy density of this halo must be rapidly increasing to
maintain the meson's velocity at $\vm<1$. However, in the
high-momentum regime, the quasiparticles and the adjoint plasma are
no longer strongly coupled to each other and the meson excitations
can achieve the speed of light. Of course, the behaviour in this
high-momentum regime restores the intuitive picture that one might
acquire from considering the theory at weak coupling, namely, that
at high momentum, the quasiparticles should be largely unaffected by
the surrounding plasma.

In the low-momentum regime, the methods which we applied in section
\ref{spectral} allowed us to estimate not only the positions of the
poles (from $\Omega$ and $\Gamma$) but also their residues $A$. The
latter were found to decrease with increasing $\gq$, as illustrated
in fig. \ref{Agrid}. As noted in \cite{asym}, the parameter
dependence of $A$ can be important in determining the overall form
of the spectral function and clearly here the fall in the residues
plays a role in the disappearance of the resonances in the spectral
functions. However, one might note these residues are indicative of
the coupling of the relevant operator to the thermal bath, \ie of
how effective the operator is in generating the relevant
quasiparticle excitations. Hence for the present purposes, we need
not think of $A$ as a physical characteristic of the quasiparticle
itself.\footnote{However, this coupling may become physically
relevant for a certain quasiparticle decay channel if in the
underlying theory the operator couples to other physical fields, \eg
photons \cite{bright}.} In contrast, $\Omega(\gq)$ and $\Gamma(\gq)$
certainly characterize the basic physical behaviour of the
quasiparticles. In particular, in order for a certain pole in the
thermal correlator to be considered a quasiparticle in the first
place, it must satisfy the Landau criterion $\Gamma\ll\Omega$.
Intuitively, we may understand this requirement by observing that
the corresponding wave function has a factor $\exp[-i\Omega t-\Gamma
t]$. Hence with $\Gamma\ll\Omega$, there are a large number of
oscillations before the excitation is damped out.

In section \ref{spectral}, we found another important reason that
the quasiparticle peaks were quickly absorbed into the background is
that the spacing between neighbouring poles shrinks with growing
momentum, as is illustrated by the dispersion curves in
fig.~\ref{omegavsqwithstrtline}. Certainly at a practical level
identifying the peaks in the spectral functions requires
$\Gamma\ll\Delta\Omega$ where $\Delta\Omega$ is the spacing between
neighbouring poles (along the real axis). As illustrated with
fig.~\ref{squash} for case VI, the peaks are begin to coalesce at
roughly the point where they vanish from the spectral function. At
this point, the poles of interest are not isolated and rather there
are a large number of poles with roughly equal spacings and
similarly growing widths, which produces some smooth continuum
(rather than individual peaks). A quasiparticle interpretation is
inappropriate then. Instead the infinite collection of poles is
collectively generating the smooth background which remains in the
spectral function.

It is interesting to consider how well the peaks of the spectral
function were reconstructed given the estimates we produced for the
relevant parameters in section \ref{spectral}. Fig.~\ref{22x} shows
a typical peak which corresponds to case IV with $\gq=50$. Also
shown is the Breit-Wigner peak \reef{peek} reconstructed with
various estimates of $\Omega$, $\Gamma$ and $A$, produced by
examining $\partial^3_\gw\gR$ as described in section
\ref{spectral}. While this reconstruction matches the shape of the
peak near the maximum quite well, there are two other notable
features which are quite apparent. First the reconstructed peak
appears displaced slightly towards smaller $\gw$ relative to where
$\partial_\gw\gR=0$. Of course, this is to be expected since there
is a rising background that also contributes to the spectral
function. The second feature, which we found surprising, is that
this background seems to be quite small. That is, the maximum of
reconstructed peak matches very well with the maximum value of the
spectral function. More generally, we found that using our parameter
estimates to reconstruct the quasiparticle peaks gave a background
that was typically only between 10 and 20\% under the top of the
peak.

\FIGURE[ht]{
\includegraphics[width=0.85 \textwidth]{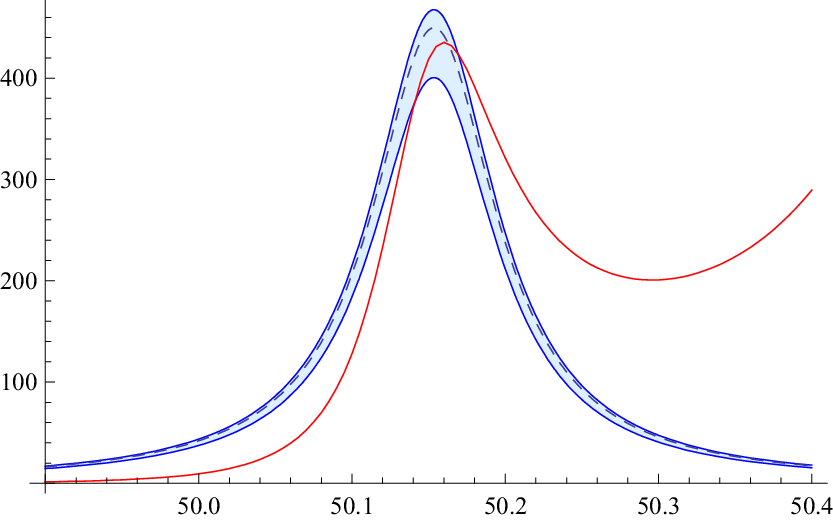}
\caption{The red curve shows the spectral function for case IV with
$\gq=50$. The blue region shows the peak \reef{peek2} as
reconstructed with the parameters $\Omega,\ \Gamma$ and $A$.
calculated in section \ref{spectral}. The center (dashed) line in
the shaded region uses $(\Omega, \Gamma, A)=(\gw_0,\bar\Gamma,
A_1)$. The bottom (solid) boundary of the shaded region is generated
with $(\Omega, \Gamma, A)=(\gw_0,\bar\Gamma+\Delta\Gamma, A_{2,+})$,
while the top (thick solid) boundary uses $(\Omega, \Gamma,
A)=(\gw_0,\bar\Gamma-\Delta\Gamma, A_{2,-})$.}\label{22x} }

Given the limitations in studying the spectral functions, it would
also be very interesting to investigate the quasinormal modes
directly, following \eg \cite{hoyos,nicknew}. While this presents
additional technical challenges, it would allow for a detailed
verification of the quasiparticle dispersion relations found here in
the low-momentum regime but also to establish the behaviour
conjectured for the high-momentum regime \cite{prep}. At this stage,
we would like to make the following observation. It is possible to
extract the quasinormal modes directly from the spectral function.
If $\gw=\Omega-i\Gamma$ is a quasinormal mode, this will show up as
a pole in the spectral function $\gR$. Since we have an approximate
idea about the location of the peaks from the spectral functions as
well as an idea about the widths, we have a reasonable initial guess
which can be fine-tuned to locate the pole. It turns out that for
larger $\gq$'s, the fine-tuning required increases. The following
table is a comparison between this method and the method of
estimating parameters from $\partial^3_\gw\gR$ for case VI.
\begin{center}
\begin{tabular}{|c|c|c|}\hline
$\gq$ & $\gw_{qn}$ & $\gw$ \\
\hline
0 & $1.031-0.042 i$ & $1.034-0.042 i$ \\
\hline
0.5 & $1.085-0.05 i$ & $1.091-0.049 i$ \\
\hline
1 & $1.227-0.072 i$ & $1.234-(0.072\pm0.001) i$ \\
\hline
1.5 & $1.408-0.112 i$ & $1.426-(0.115\pm 0.001)i $ \\
\hline
2 & $1.596-0.164 i$ & $1.644-(0.186\pm0.004)i$ \\
\hline
2.2 & $1.671-0.187 i$ & $1.730-(0.220\pm 0.010)i$ \\
\hline
\end{tabular}
\end{center}
As is clear from the table, for larger $\gq$ the quasinormal mode
analysis predicts a lower $\Gamma$.

While we see a dramatic rise in the width $\Gamma$ of quasiparticles
in the present model, it remains to understand if this increase is a
universal feature which emerges in any holographic model. Of course,
an even more important question, is whether or not such an effect is
realized in the strongly coupled quark-gluon plasma of QCD.
Certainly both of these questions deserve further investigation.

At present, investigating spectral functions with non-zero momentum
is an important direction of ongoing research \cite{hotp,hotp2}.
Only partial results are available since the existing methods seem
to be inadequate for this problem. Recent results, which generally
use the so-called potential approach, indicate the spectral
functions flatten with increasing momentum \cite{hotp,hotp2} -- an
effect consistent with our present findings.

Of course, if these effects found here in our holographic studies
are realized in QCD, they would have interesting implications for
experiments at RHIC and LHC. In particular, as suggested in
\cite{ejaz,wind}, they could lead to a significant additional
suppression of $J/\Psi$ or other heavy quark mesons with large
transverse momenta. Certainly a critical momentum where the
quasiparticle widths rise dramatically would produce a dramatic
effect. Another potentially interesting effect was outlined in
\cite{bright}. One distinct feature resulting from a limiting
velocity $\vm<1$ is that the four momenta of quasiparticles become
null at some point. For example, if we drop the higher order terms
in the asymptotic dispersion relation \reef{strait}, then the
quasiparticle's dispersion relation crosses the null cone at
 \be \Omega={a_i\over1-\vm}=\gq\,. \label{intersect} \ee
Following the holographic techniques of \cite{first}, one then finds
that this crossing of the null cone produces a peak in photon
production from charged quasiparticles \cite{bright}. Such a
resonance would then be a distinctive experimental signature of
$\vm<1$. Of course, as speculated in fig. \ref{possibility}, the
dispersion relations may cross the null cone twice and exhibit a
regime $\partial_\gq\Omega>1$, which may produce further dramatic
signals.

From \reef{intersect}, we observe that this enhancement is pushed to
infinite momentum as $\vm$ approaches one. Hence, if $\vm$ is still
close to one at RHIC or LHC, this signal would only appear at very
large momenta. Further, if the quasiparticle width increases too
quickly, this enhancement in the photon production would likely be
washed out. The point where the quasiparticle `four-momentum' is
null is indicated in fig.~\ref{Gammavsqfromplots} for the four cases
in table \reef{table2}. For cases IV, V and VI, one indeed finds
that this null momentum seems to lie in the regime where the width
is growing rapidly, while case VII seems to present an exception to
this rule. Further as can be seen in fig.~\ref{omegaslope} for the
more stable cases presented in table \reef{table}, the quasiparticle
dispersion relations seem to cross the null cone well away from the
maximum momentum. Hence it seems this question also requires further
study.


\acknowledgments It is a pleasure to thank Andreas Karch, Hong Liu,
Juan Maldacena, Amanda Peet, Peter Petreczky, Krishna Rajagopal,
Lennie Susskind and especially Andrei Starinets for useful
correspondence and conversations. Research at Perimeter Institute is
supported by the Government of Canada through Industry Canada and by
the Province of Ontario through the Ministry of Research \&
Innovation. RCM also acknowledges support from an NSERC Discovery
grant and funding from the Canadian Institute for Advanced Research.
RCM thanks the organizers of {\it Miami 2007} and the {\it
LindeFest} for the opportunity to speak on this work. AS also thanks
the organizers of {\it Quark Matter 2008} for the opportunity to
speak on this work.

\appendix
\section{WKB approximation} \label{wonk}

In this section, we outline how WKB calculations might be applied to
produce an approximate value for the quasinormal eigenfrequencies.
Recall that these eigenfrequencies are determined by solving the
relevant wave equation with boundary conditions of an ingoing wave
at the horizon and of only the normalizable mode asymptotically
\cite{hoyos}. In \reef{schrod}, the relevant wave equation for the
transverse vector modes has been caste into the form of a
one-dimensional Schr\"odinger with an effective energy
 \begin{eqnarray}\label{effect}
E_{eff}&=&\Omega_S-i\Gamma_S\\
&=&\gw^2=\Omega^2-\Gamma^2-2i\,\Omega\Gamma\,. \nonumber
 \end{eqnarray}
The quasinormal eigenfrequencies correspond to the positions of the
poles in the thermal correlators and so we denote the eigenfrequency
$\gw=\Omega - i\Gamma$ above in keeping with the notation of section
\ref{spectral}.

In a typical case of interest, the effective potential has a form as
illustrated in figs. \ref{10x} or \ref{12x}, with a well that is
separated from the horizon at $\rho=1$ by a large potential barrier.
The relevant modes corresponding to metastable meson states are then
bound states with support primarily in the potential well but which
slowly tunnel out through the barrier. The idea then is to
approximately determine $E_{eff}$ for these modes using WKB
techniques -- see, \eg \cite{qbook}.

First the real part of the effective energy $\Omega_S$ is obtained
by fine-tuning $\Omega_S$ so that
 \be\label{WKBE}
\int_{R_2}^{R_3}dR_*
\sqrt{\Omega_S-V(R_*)}=\left(n-{1\over2}\right)\,\pi\,,
 \ee
where $R_2,R_3$ are the classical turning points in the potential
well where $V=\Omega_S$. The first peak in one of the spectral
functions would correspond to $n=1$. Next within the WKB
approximation, the decay rate $\Gamma_S$ is defined as \cite{qbook}
 \be\label{WKBG}
\Gamma_S={\cal N}\exp(-2\int_{R_1}^{R_2}dR_*\sqrt{V-\Omega_S})\,.
 \ee
Here $R_1,R_2$ denote classical turning points in the barrier and
implicitly $\Omega_S$ denotes the WKB energy eigenvalue determined
by \reef{WKBE}. The normalization constant ${\cal N}$ is given by
 \be {\cal N}^{-1}=\int_{R_2}^{R_3} {dR_*\over \sqrt{V-\Omega_S}}\,.
 \ee

Given these WKB results for the Schr\"odinger problem, we may use
\reef{effect} to determine the eigenfrequency for the corresponding
quasinormal mode:
 \ba\label{result}
\Omega^2&=&{\Omega_S\over2}\left[1+\sqrt{1+\Gamma_S^2/\Omega_S^2}\right]
\simeq \Omega_S\left[1+{1\over 4}{\Gamma_S^2\over\Omega_S^2}\right]\,,\\
\Gamma&=&\Gamma_S/2\Omega\simeq {\Gamma_S\over 2\sqrt{\Omega_S}}
\left[1-{1\over 8}{\Gamma_S^2\over\Omega_S^2}\right]\,.\nonumber
 \ea
In the second of each of these expressions, we have presented the
leading terms in an expansion with $\Gamma_S/\Omega_S\ll1$, which is
a necessary condition for the validity of the WKB approximation. The
latter calculation focussed on the well and barrier appearing in the
effective potential $V(R_*)$ while completely ignoring the horizon.
In the results of, \eg \cite{spectre,hoyos,quasi}, one sees that for
an effective potential is a monotonically rising function
$\Gamma\sim\Omega$ or $\Gamma_S>\Omega_S$. Clearly, the horizon has
an important effect in determining the eigenfrequencies and, in
particular, the decay width in this situation. Hence one cannot
expect that the WKB calculations outlined above will be reliable in
this regime. In particular, when the barrier is getting small, the
WKB calculation would yield an tunnelling rate that becomes large
but the precise value of $\Gamma_S$ would depend crucially on the
structure of the potential near and the boundary conditions at the
horizon.

However, as discussed in section \ref{quasi}, there are additional
subtleties in the WKB approximation. For the WKB calculation of the
tunnelling rate to be valid, the change of the momentum over a
wavelength must be small compared to the momentum itself. That is,
if we define $k=\sqrt{V-\gw^2}$ under the barrier, then we must
require \cite{qbook},
 \be\label{condo}
{|k'(R_*)|\over k(R_*)^2}\ll 1\,.
 \ee

\FIGURE[ht]{
\includegraphics[width=0.7 \textwidth]{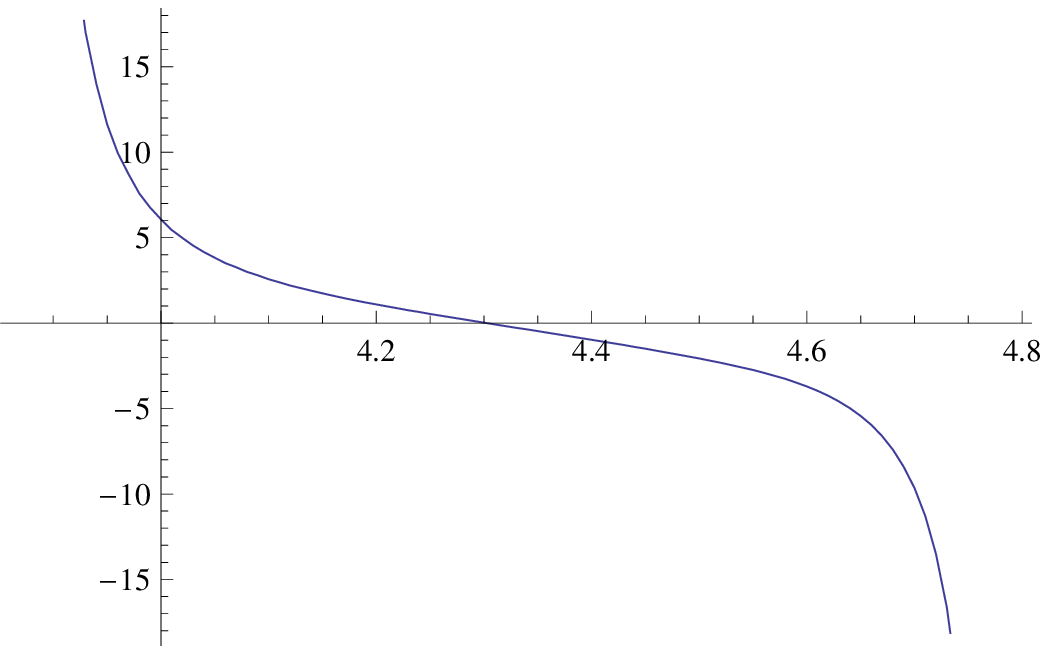}
\caption{The validity of the WKB approximation can be tested by
considering $k'/k^2$ versus $\rho$. These are plotted here inside
the barrier for $\tilde d=0.25,\chi_0=0.99$. At the turning points a
deviation from unity is expected but significant deviations are seen
well away from those points, making the WKB approximation
\reef{WKBG} unreliable.} \label{WKB}}
Unfortunately, in the examples considered in the paper, this
condition is typically not satisfied for a large region within the
barrier, essentially because the shoulders of the barrier are too
steep -- \eg see fig.~\ref{WKB}. This makes the WKB calculation
\reef{WKBG} of $\Gamma_S$ unreliable, in particular in the regime
where $\Gamma$ begins to increase dramatically. We have done WKB
calculations of $\Gamma$ in various cases when
$\Gamma/\Gamma_0\sim1$ and found the WKB results agreed with those
in section \ref{spectral} to within roughly 15\% despite the issues
discussed above. Note that the condition \reef{condo} is quite
generally satisfied for the calculation \reef{WKBE} of $\Omega_S$,
the real part of the effective energy. Hence in the regime
$\Gamma_S/\Omega_S\ll1$, the WKB approach would still provide an
accurate approximation for $\Omega\simeq\Omega_S^{1/2}$. Further,
one can still gain some qualitative insight into $\Gamma$ by
considering when the barrier disappears in the effective potential,
as discussed in section \ref{quasi}.


\begin{thebibliography}{99}

\bibitem
{juan}
J.M.~Maldacena, ``The large N limit of superconformal field
theories and supergravity,'' Adv.\ Theor.\ Math.\ Phys.\  {\bf 2}
 231 (1998) [Int.\ J.\ Theor.\ Phys.\  {\bf 38}  1113 (1999)]
[arXiv:hep-th/9711200].

\bibitem
{bigRev}
  O.~Aharony, S.~S.~Gubser, J.~M.~Maldacena, H.~Ooguri and Y.~Oz,
  ``Large N field theories, string theory and gravity,''
  Phys.\ Rept.\  {\bf 323}, 183 (2000)
  [arXiv:hep-th/9905111].


\bibitem
{shear}
P.~Kovtun, D.T.~Son and A.O.~Starinets, ``Viscosity in
strongly interacting quantum field theories from black hole
  physics,''
  Phys.\ Rev.\ Lett.\  {\bf 94}, 111601 (2005)
  [arXiv:hep-th/0405231];\\
P.~Kovtun, D.T.~Son and A.O.~Starinets, ``Holography and
hydrodynamics: Diffusion on stretched horizons,''
  JHEP {\bf 0310}, 064 (2003)
  [arXiv:hep-th/0309213].


\bibitem
{seat}
C.P.~Herzog, A.~Karch, P.~Kovtun, C.~Kozcaz and L.G.~Yaffe,
``Energy loss of a heavy quark moving through N = 4 supersymmetric
Yang-Mills plasma,'' JHEP {\bf 0607}, 013 (2006)
  [arXiv:hep-th/0605158];\\
H.~Liu, K.~Rajagopal and U.A.~Wiedemann, ``Calculating the jet
quenching parameter from AdS/CFT,''
  Phys.\ Rev.\ Lett.\  {\bf 97}  182301 (2006)
  [arXiv:hep-ph/0605178];\\
S.S.~Gubser, ``Drag force in AdS/CFT,''
  Phys.\ Rev.\  D {\bf 74}  126005 (2006)
  [arXiv:hep-th/0605182];\\
J.~Casalderrey-Solana and D.~Teaney, ``Heavy quark diffusion in
strongly coupled N = 4 Yang Mills,''
  Phys.\ Rev.\  D {\bf 74}, 085012 (2006)
  [arXiv:hep-ph/0605199].

\bibitem
{johanna}
J.~Babington, J.~Erdmenger, N.J.~Evans, Z.~Guralnik and
I.~Kirsch, ``Chiral symmetry breaking and pions in
non-supersymmetric gauge/gravity duals,'' Phys.\ Rev.\ D {\bf 69}
 066007 (2004)
[arXiv:hep-th/0306018];\\
M.~Kruczenski, D.~Mateos, R.C.~Myers and D.J.~Winters,
  ``Towards a holographic dual of large-$\nc$ QCD,''
  JHEP {\bf 0405}, 041 (2004)
  [arXiv:hep-th/0311270]; \\
I.~Kirsch, ``Generalizations of the AdS/CFT correspondence,''
Fortsch.\ Phys.\  {\bf 52}  727 (2004) [arXiv:hep-th/0406274].\\

\bibitem
{prl}
D.~Mateos, R.C.~Myers and R.M.~Thomson, ``Holographic phase
transitions with fundamental matter,''
  Phys.\ Rev.\ Lett.\  {\bf 97}  091601 (2006)
  [arXiv:hep-th/0605046].

\bibitem
{long}
D.~Mateos, R.C.~Myers and R.M.~Thomson, ``Thermodynamics of
the brane,''
  JHEP {\bf 0705}, 067 (2007)
  [arXiv:hep-th/0701132].

\bibitem
{recent}
D.~Mateos, R.C.~Myers and R.M.~Thomson, ``Holographic
viscosity of fundamental matter,'' Phys. Rev. Lett. {\bf 98}
101601 (2007) [arXiv:hep-th/0610184];\\
T.~Albash, V.~Filev, C.V.~Johnson and A.~Kundu, ``A
topology-changing phase transition and the dynamics of flavour,''
arXiv:hep-th/0605088;\\
A.~Karch and A.~O'Bannon,
  ``Chiral transition of N = 4 super Yang-Mills with flavor on a 3-sphere,''
  Phys.\ Rev.\  D {\bf 74}  085033 (2006)
  [arXiv:hep-th/0605120];\\
T.~Albash, V.~Filev, C.V.~Johnson and A.~Kundu, ``Global currents,
phase transitions, and chiral symmetry breaking in large $\nc$ gauge
theory,'' arXiv:hep-th/0605175;\\
A.~Karch and A.~O'Bannon,
  ``Metallic AdS/CFT,''
  arXiv:0705.3870 [hep-th];\\
A.~O'Bannon, ``Hall Conductivity of Flavor Fields from AdS/CFT,''
  arXiv:0708.1994 [hep-th];\\
J.~Erdmenger, N.~Evans, I.~Kirsch and E.~Threlfall, ``Mesons in
Gauge/Gravity Duals - A Review,''
  arXiv:0711.4467 [hep-th].

\bibitem
{spectre} R.C.~Myers, A.O.~Starinets and R.M.~Thomson, ``Holographic
spectral functions and diffusion constants for fundamental matter,''
  JHEP {\bf 0711}, 091 (2007)
  [arXiv:0706.0162 [hep-th]].

\bibitem
{hoyos} C.~Hoyos, K.~Landsteiner, and S.~Montero, ``Holographic
meson melting,'' JHEP {\bf 0704} 031  (2007) [arXiv:hep-th/0612169].

\bibitem
{lattice} T.~Umeda, K.~Nomura and H.~Matsufuru, ``Charmonium at
finite temperature in quenched lattice QCD,''   Eur.\ Phys.\ J.\  C
{\bf 39S1} 9  (2005)
  [arXiv:hep-lat/0211003];\\
  %
  M.~Asakawa and T.~Hatsuda,
  ``J/$\psi$ and $\eta$/c in the deconfined plasma from lattice QCD,''
   Phys.\ Rev.\ Lett.\  {\bf 92}  012001 (2004)
  [arXiv:hep-lat/0308034];\\
%
  S.~Datta, F.~Karsch, P.~Petreczky and I.~Wetzorke,
  ``Behavior of charmonium systems after deconfinement,''
  Phys.\ Rev.\  D {\bf 69}  094507 (2004)
  [arXiv:hep-lat/0312037];\\
%
A.~Jakovac, P.~Petreczky, K.~Petrov and A.~Velytsky, ``On charmonia
survival above deconfinement,''   [arXiv:hep-lat/0603005];\\
%
  G.~Aarts, C.R.~Allton, R.~Morrin, A.P.O.~Cais, M.B.~Oktay,
  M.J.~Peardon and J.I.~Skullerud,
  ``Charmonium spectral functions in $N_{f}$ = 2 QCD at high temperature,''
  PoS {\bf LAT2006}  126 (2006)
  [arXiv:hep-lat/0610065];\\
%
G.~Aarts, C.~Allton, J.~Foley, S.~Hands and S.~Kim,
  ``Spectral functions at nonzero momentum in hot QCD,''
  PoS {\bf LAT2006} 134  (2006)
  [arXiv:hep-lat/0610061];\\
%
  P.~Petreczky,
  ``Lattice QCD at finite temperature,''
  Nucl.\ Phys.\  A {\bf 785}  10 (2007)
  [arXiv:hep-lat/0609040];\\
%
   G.~Aarts, C.~Allton, J.~Foley, S.~Hands and S.~Kim,
  ``Spectral functions at small energies and the electrical conductivity in
  hot, quenched lattice QCD,''
  Phys.\ Rev.\ Lett.\  {\bf 99}  022002 (2007)
  [arXiv:hep-lat/0703008];\\
%
  H.B.~Meyer,
  ``A calculation of the shear viscosity in SU(3) gluodynamics,''
  arXiv:0704.1801 [hep-lat]; \\
  %
  G.~Aarts, C.~Allton, M.~B.~Oktay, M.~Peardon and J.~I.~Skullerud,
  ``Charmonium at high temperature in two-flavor QCD,''
  arXiv:0705.2198 [hep-lat].
%
H.~Satz, ``Colour deconfinement and quarkonium binding,''
  J.\ Phys.\ G {\bf 32}, R25 (2006)
  [arXiv:hep-ph/0512217].

\bibitem
{ejaz} Q.J.~Ejaz, T.~Faulkner, H.~Liu, K.~Rajagopal and
U.A.~Wiedemann,
  ``A limiting velocity for quarkonium propagation in a strongly coupled plasma
  via AdS/CFT,''
  arXiv:0712.0590 [hep-th].

\bibitem
{findens} S.~Kobayashi, D.~Mateos, S.~Matsuura, R.C.~Myers and
R.M.~Thomson, ``Holographic phase transitions at finite baryon
density,''
  JHEP {\bf 0702}, 016 (2007)
  [arXiv:hep-th/0611099].

\bibitem
{finitemu} D.~Mateos, S.~Matsuura, R.C.~Myers and R.M.~Thomson,
``Holographic phase transitions at finite chemical potential,''
  JHEP {\bf 0711}, 085 (2007)
  [arXiv:0709.1225 [hep-th]].

\bibitem
{johanna2} J.~Erdmenger, M.~Kaminski and F.~Rust, ``Holographic
vector mesons from spectral functions at finite baryon or isospin
density,''
  arXiv:0710.0334 [hep-th].

\bibitem{iancu}
  Y.~Hatta, E.~Iancu and A.~H.~Mueller,
  ``Deep inelastic scattering off a N=4 SYM plasma at strong coupling,''
  JHEP {\bf 0801}, 063 (2008)
  [arXiv:0710.5297 [hep-th]];\\
  Y.~Hatta, E.~Iancu and A.~H.~Mueller,
  ``Jet evolution in the N=4 SYM plasma at strong coupling,''
  arXiv:0803.2481 [hep-th].



\bibitem
{flavour} O.~Aharony, A.~Fayyazuddin and J.~M.~Maldacena, ``The
large N limit of N = 2,1 field theories from three-branes in
F-theory,'' JHEP {\bf 9807}, 013 (1998)
  [arXiv:hep-th/9806159];\\
A.~Karch and L.~Randall, ``Open and closed string interpretation of
SUSY CFT's on branes with boundaries,'' JHEP {\bf
0106}  063 (2001) [arXiv:hep-th/0105132];\\
A.~Karch and E.~Katz, ``Adding flavor to AdS/CFT,'' JHEP {\bf 0206}
043 (2002) [arXiv:hep-th/0205236].

\bibitem
{us-meson} M.~Kruczenski, D.~Mateos, R.C.~Myers and D.J.~Winters,
``Meson spectroscopy in AdS/CFT with flavour,'' JHEP {\bf 0307}
 049 (2003) [arXiv:hep-th/0304032].

\bibitem
{zero} A.~Karch and A.~O'Bannon, ``Holographic Thermodynamics at
Finite Baryon Density: Some Exact Results,'' JHEP {\bf 0711}, 074
(2007) [arXiv:0709.0570 [hep-th]].

\bibitem
{new} D.T.~Son and A.O.~Starinets, ``Viscosity, Black Holes, and
Quantum Field Theory,'' arXiv:0704.0240 [hep-th].

\bibitem
{quasi} P.K.~Kovtun and A.O.~Starinets, ``Quasinormal modes and
holography,'' Phys.\ Rev.\  D {\bf 72}, 086009 (2005)
  [arXiv:hep-th/0506184].

\bibitem
{Son:2002sd} D.T.~Son and A.O.~Starinets,
   ``Minkowski-space correlators in AdS/CFT correspondence: Recipe and
  applications,''
JHEP {\bf 0209}, 042 (2002)
  [arXiv:hep-th/0205051].

\bibitem
{techn} D.~Teaney, ``Finite temperature spectral densities of
momentum and R-charge  correlators in N = 4 Yang Mills theory,''
   Phys.\ Rev.\  D {\bf 74}, 045025 (2006) [arXiv:hep-ph/0602044];\\
P.~Kovtun and A.O.~Starinets, ``Thermal spectral functions of
strongly coupled N = 4 supersymmetric
  Yang-Mills theory,''
  Phys.\ Rev.\ Lett.\  {\bf 96}, 131601 (2006)
  [arXiv:hep-th/0602059].

\bibitem
{Birmingham:2001pj}
  D.~Birmingham, I.~Sachs and S.~N.~Solodukhin,
  ``Conformal field theory interpretation of black hole quasi-normal modes,''
 Phys.\ Rev.\ Lett.\  {\bf 88}, 151301 (2002)
  [arXiv:hep-th/0112055].

\bibitem
{prep} R.C.~Myers and A.~Sinha, in preparation.



\bibitem
{nicknew} N.~Evans and E.~Threlfall,
  ``Mesonic quasinormal modes of the Sakai-Sugimoto model at high
  temperature,''
  arXiv:0802.0775 [hep-th].


\bibitem
{andrei1} A.~Nunez and A.O.~Starinets, ``AdS/CFT correspondence,
quasinormal modes, and thermal correlators in N  = 4 SYM,''
  Phys.\ Rev.\  D {\bf 67}  124013 (2003)
  [arXiv:hep-th/0302026].

\bibitem
{andrei2} A.O.~Starinets, ``Quasinormal modes of near extremal black
branes,''
  Phys.\ Rev.\  D {\bf 66} 124013  (2002)
  [arXiv:hep-th/0207133].

\bibitem
{gong} A.~Paredes, K.~Peeters and M.~Zamaklar, ``Mesons versus
quasi-normal modes: undercooling and overheating,''
  arXiv:0803.0759 [hep-th].


\bibitem
{qbook} See, for example: E.~Merzbacher, ``Quantum Mechanics,'' John
Wiley \& Sons, 1967.

\bibitem
{superfast} R.~Fox, C.G.~Kuper and S.G.~Lipson, ``Faster-than-light
group velocities and causality violation,''
  Proc.\ Roy.\ Soc.\ Lond.\  {\bf 316}, 515 (1970);
``Do faster-than-light group velocities imply violation of
causality?,''
  Nature {\bf 223}, 597 (1969).

\bibitem
{flurry} S.S.~Gubser, ``Momentum fluctuations of heavy quarks in the
gauge-string duality,''
  Nucl.\ Phys.\  B {\bf 790}, 175 (2008)
  [arXiv:hep-th/0612143];\\
H.~Liu, K.~Rajagopal and U.A.~Wiedemann, ``Wilson loops in heavy ion
collisions and their calculation in AdS/CFT,''
  JHEP {\bf 0703}, 066 (2007)
  [arXiv:hep-ph/0612168];\\
P.C.~Argyres, M.~Edalati and J.F.~Vazquez-Poritz, ``Spacelike
strings and jet quenching from a Wilson loop,''
  JHEP {\bf 0704}, 049 (2007)
  [arXiv:hep-th/0612157];\\
J.~Casalderrey-Solana and D.~Teaney, ``Transverse momentum
broadening of a fast quark in a N = 4 Yang Mills plasma,''
  JHEP {\bf 0704}, 039 (2007)
  [arXiv:hep-th/0701123].

\bibitem
{wind} H.~Liu, K.~Rajagopal and U.~A.~Wiedemann,
  ``An AdS/CFT calculation of screening in a hot wind,''
  Phys.\ Rev.\ Lett.\  {\bf 98}, 182301 (2007)
  [arXiv:hep-ph/0607062];\\
K.~Peeters, J.~Sonnenschein and M.~Zamaklar, ``Holographic melting
and related properties of mesons in a quark gluon plasma,''
  Phys.\ Rev.\  D {\bf 74}, 106008 (2006)
  [arXiv:hep-th/0606195];\\
M.~Chernicoff, J.A.~Garcia and A.~Guijosa, ``The energy of a moving
quark-antiquark pair in an N = 4 SYM plasma,''
  JHEP {\bf 0609}, 068 (2006)
  [arXiv:hep-th/0607089].




\bibitem
{news} S.S.~Gubser and A.~Yarom, ``Linearized hydrodynamics from
probe-sources in the gauge-string duality,''
  arXiv:0803.0081 [hep-th].


\bibitem
{asym} I.~Amado, C.~Hoyos-Badajoz, K.~Landsteiner and S.~Montero,
``Residues of Correlators in the Strongly Coupled N=4 Plasma,''
  arXiv:0710.4458 [hep-th].

\bibitem
{bright} D.~Mateos and L.~Patino, ``Bright branes for strongly
coupled plasmas,''
  JHEP {\bf 0711}, 025 (2007)
  [arXiv:0709.2168 [hep-th]].

\bibitem
{hotp} S.~Datta, F.~Karsch, S.~Wissel, P.~Petreczky and I.~Wetzorke,
  ``Charmonia at finite momenta in a deconfined plasma,''
  arXiv:hep-lat/0409147.
G.~Aarts, C.~Allton, J.~Foley, S.~Hands and S.~Kim,
  ``Spectral functions at non-zero momentum in hot QCD,''
  PoS {\bf LAT2006}, 134 (2006)
  [arXiv:hep-lat/0610061];\\
G.~Aarts, C.~Allton, J.~Foley, S.~Hands and S.~Kim,
  ``Meson spectral functions at nonzero momentum in hot QCD,''
  Nucl.\ Phys.\  A {\bf 785}, 202 (2007)
  [arXiv:hep-lat/0607012];\\
G.~Aarts, S.~Hands, S.Y.~Kim and J.M.~Martinez Resco,
  ``On meson spectral functions at high temperature and nonzero momentum,''
  PoS {\bf LAT2005}, 182 (2006)
  [arXiv:hep-lat/0509062];\\
W.M.~Alberico, A.~Beraudo, P.~Czerski and A.~Molinari,
  ``Finite momentum meson correlation functions in a QCD plasma,''
  Nucl.\ Phys.\  A {\bf 775}, 188 (2006)
  [arXiv:hep-ph/0605060];\\
W.M.~Alberico, A.~Beraudo, A.~Czerska, P.~Czerski and A.~Molinari,
  ``Meson screening masses in the interacting QCD plasma,''
  Nucl.\ Phys.\  A {\bf 792}, 152 (2007)
  [arXiv:hep-ph/0703298];\\

\bibitem
{hotp2} P. Petreczky, private communication.



\bibitem
{first} S.~Caron-Huot, P.~Kovtun, G.D.~Moore, A.~Starinets and
L.G.~Yaffe, ``Photon and dilepton production in supersymmetric
Yang-Mills plasma,''
  JHEP {\bf 0612}, 015 (2006)
  [arXiv:hep-th/0607237].





\end{thebibliography}
\end{document}